\documentclass[oneside,leqno,english]{amsart}

\usepackage[T2A,T1]{fontenc}
\usepackage[latin9]{inputenc}
\usepackage{geometry}
\usepackage{mathrsfs}
\usepackage{amsbsy}
\usepackage{amstext}
\usepackage{amsthm}
\usepackage{amssymb}
\usepackage{graphicx}
\usepackage[authoryear]{natbib}

\makeatletter

\DeclareRobustCommand{\cyrtext}{%
  \fontencoding{T2A}\selectfont\def\encodingdefault{T2A}}
\DeclareRobustCommand{\textcyr}[1]{\leavevmode{\cyrtext #1}}

\numberwithin{equation}{section}
\numberwithin{figure}{section}
\numberwithin{table}{section}
\theoremstyle{definition}
\newtheorem*{defn*}{\protect\definitionname}
\theoremstyle{plain}
\newtheorem*{prop*}{\protect\propositionname}
\theoremstyle{plain}
\newtheorem*{cor*}{\protect\corollaryname}
\theoremstyle{plain}
\newtheorem*{thm*}{\protect\theoremname}
\newenvironment{lyxlist}[1]
	{\begin{list}{}
		{\settowidth{\labelwidth}{#1}
		 \setlength{\leftmargin}{\labelwidth}
		 \addtolength{\leftmargin}{\labelsep}
		 }}
	{\end{list}}
\theoremstyle{plain}
\newtheorem*{lem*}{\protect\lemmaname}
\theoremstyle{remark}
\newtheorem*{rem*}{\protect\remarkname}
\theoremstyle{remark}
\newtheorem*{note*}{\protect\notename}

\@ifundefined{date}{}{\date{}}
\usepackage{stix}
\usepackage{amsaddr}

\allowdisplaybreaks
\setcitestyle{citesep={,}}

\DeclareMathOperator*{\ord}{ord}

\makeatother

\usepackage{babel}
\providecommand{\corollaryname}{Corollary}
\providecommand{\definitionname}{Definition}
\providecommand{\lemmaname}{Lemma}
\providecommand{\notename}{Note}
\providecommand{\propositionname}{Proposition}
\providecommand{\remarkname}{Remark}
\providecommand{\theoremname}{Theorem}

\begin{document}
\begin{center}
{\footnotesize{}SIAM J. Control and Optimization, vol. 19, No. 4,
pp. 445-468, 1981}.\\
Reprints available from hammer@mst.ufl.edu
\par\end{center}
\title[]{CAUSAL FACTORIZATION AND LINEAR FEEDBACK$\dagger$}
\thanks{$\dagger$The latex version of this paper was obtained from the original
by using optical character recognition at Mathpix.com. We express
our profound thanks to Mathpix.}
\author{Jacob Hammer {\footnotesize and} Michael Heymann}
\begin{abstract}
Abstract. An algebraic framework for the investigation of linear dynamic
output feedback is introduced. Pivotal in the present theory is the
problem of causal factorization, i.e. the problem of factoring two
systems over each other through a causal factor. The basic issues
are resolved with the aid of the new concept of latency kernels.
\end{abstract}

\maketitle

\section*{1. Introduction.}

In recent years the system theory literature has seen a rapidly growing
interest in questions associated with linear feedback. In the early
1960 's, linear control theory centered chiefly around quadratic (Gaussian)
optimal problems and the resulting feedback designs. Later, interest
in feedback shifted to a variety of so-called "synthesis" problems.
These included the well-known problem of observer design (see Luenberger
{[}1966{]} ), the pole shifting theorem and related issues (Wonham
{[}1967{]} Simon and Mitter {[}1968{]}, Brash and Pearson {[}1970{]},
Heymann {[}1968{]}) as well as the decoupling problem (Falb and Wolovich
{[}1967{]}, Gilbert {[}1969{]}, Wonham and Morse $[1970],$ Morse
and Wonham {[}1970{]} ). All of these feedback synthesis problems,
as well as many others, were formulated and resolved within the framework
of state space representations. While most of the work was done with
the use of conventional state equations, the work of Wonham and Morse
was distinguished by its "coordinate free" setting and initiated
what later developed into the celebrated "geometric theory" of linear
control (see, e.g., Wonham {[}1979{]}).

The current growing interest in linear feedback differs significantly
from that of the past both in character and in its source of motivation.
While previously the study of feedback was largely oriented at problem
solving, the current interest is motivated by a desire of gaining
insight into the general nature of linear feedback-chiefly from an
algebraic point of view. Much of the motivation for the present trend
can be traced back to the work of Rosenbrock {[}1970{]} , in which
polynomial matrix techniques were used for the study of a variety
of (linear) control theoretic questions. Particularly useful turned
out to be techniques based on polynomial fraction representations
of transfer functions (see, e.g., Heymann {[}1972{]}, Wolovich {[}1974{]},
Forney {[}1975{]}, Fuhrmann {[}1976{]}). In this setting of fraction
representations, feedback was first studied in Heymann {[}1972{]}
(see especially Chapter 6 therein), and in a polynomial module framework
the study of feedback was initiated by Eckberg {[}1974{]}. State feedback
also received attention in an algebraic framework by Morse {[}1975{]}.
A different approach to the study of linear feedback was taken in
Hautus and Heymann {[}1978{]}, where the fundamental underlying object
was taken to be the input-output map of the system. There, static
linear state feedback was investigated in an algebraic framework consistent
with the setting of the (classical) module theory of linear realization
as introduced by Kalman (see, e.g., Kalman et al. $[1969,\text{ Chapter }10]$
). More recently, state feedback was also examined in Fuhrmann {[}1979{]}
using what he termed "polynomial models", and in Münzner and Prätzel-Wolters
(1979a{]}, {[}1979b{]}, {[}1979c{]} in a module and category theoretic
framework.

While these various approaches to the study of feedback differ from
each other substantially both in the underlying concept and in philosophy,
they commonly converge on essentially the same (standard) issues that
characterize state feedback. It is significant, however, that no success
(and, in fact, very little effort, if any) has been reported in respect
to output, as opposed to state feedback. When various fundamental
questions in regard to output feedback are examined, it becomes immediately
clear that difficulties arise that are completely absent in the state-feedback
setting. In fact, one discovers immediately that crucial insight is
missing. It turns out that the chief reason for this state of affairs
is the fact that all of the presently existing algebraic theory of
linear systems, and especially that of feedback, rests in one way
or another on the theory of modules over the ring $K[z]$ of polynomials
and on polynomial matrices. This algebraic machinery is completely
satisfactory to develop a fairly comprehensive framework for state
feedback. It is not adequate, though, to deal with output-feedback
where issues associated with causality become significantly more intricate.

The present paper deals in a comprehensive way with the problem of
causal output feedback. A related question which receives a great
deal of attention in the paper and on which much of the theory hinges
is the so-called causal factorization problem. This is the problem
of when a given linear input-output map can be factored over another
one by a causal linear map. Through the resolution of this issue,
questions associated with dynamic causal output feedback are then
also resolved. Attention is also given to the static factorization
problem as well as the problem of static feedback where special emphasis
is placed on the state-feedback case.

A crucial role in the present theory is played by the newly introduced
concept of latency. In the discrete time setting, latency expresses
"degree of causality" and (intuitively) refers to the intrinsic
delay which inputs encounter before output responses are produced.
Latency is algebraically expressed by modules over the ring $K\left[\left[z^{-1}\right]\right]$
of power series (in $z^{-1}$ over a field $K$). These modules arise
in a natural way when the concept of causality is studied algebraically
and in fact are readily seen to be the natural algebraic device for
the study of feedback.

The paper is organized as follows. In $\S2$ the basic concepts of
$\Lambda K$ -linear maps causality, linear i/o maps as well as linear
i/s maps, which have been investigated in detail in Hautus and Heymann
{[}1978{]}, are reviewed. The conceptual viewpoint, on which the present
investigation of feedback rests, is discussed in $\S3.$ An important
technical concept that arises in the algebraic study of linear systems
both in connection with the $K[z]$ -module theory and the $K\left[\left[z^{-1}\right]\right]$
-module theory is that of "proper bases" and "proper independence".
This is the topic of $\$4.$ Section 5 is devoted to the investigation
of causal factorization, the main result being Theorem 5.2 and its
corollaries. Results are also obtained on static feedback (Theorems
5.10 and 5.14 ). In $\S6$ the problem of invariants is investigated
in detail and explicit characterizations are derived and exhibited.
The role of the latency kernels and latency indices is also discussed.
The paper is concluded in $\$7$ with an investigation of the interesting
question of feedback (design) limitations. It is shown that the essential
limitation to the possibility of causal feedback implementation of
precompensators is the system's latency. In particular, precompensators
can be implemented as causal feedback devices modulo a "precompensator
remainder" whose dynamic order need not exceed the sum of the system's
latency indices.

\section*{2. $\Lambda\mathrm{K}$ -linear maps, causality and input-output
behavior.}

We shall adopt a terminology and setup consistent with that of Hautus
and Heymann {[}1978{]}.

Let $K$ be a field and let $S$ be a $K$ -linear space. The class
of all truncated $S$ -valued Laurent series of the form
\[
(2.1)\quad s=\sum_{t=t_{0}}^{\infty}s_{t}z^{-t}
\]
is denoted by $S\left(\left(z^{-1}\right)\right)$ or alternatively
by $\Lambda S$. The polynomial subset of $S$, i.e., the set of all
elements of $\Lambda S$ of the form $\sum_{t\leq0}s_{t}z^{-t},$
is denoted $\Omega^{+}S.$ The power series subset of $\Lambda S$
i.e., the set of all elements of the form $\sum_{t\geq0}s_{t}z^{-t},$
is denoted $\Omega^{-}S.$ The set $\Lambda K=K\left(\left(z^{-1}\right)\right)$
of $K$ -valued Laurent series is endowed with a field structure under
the operation of convolution as multiplication and coefficientwise
addition. In particular, for $\alpha=$ $\sum_{t=t_{0}}^{\infty}\alpha_{t}z^{-t}$
and $\alpha^{\prime}=\sum_{t=t_{i}}^{\infty}\alpha_{t}^{\prime}z^{-t}$
in $\Lambda K,$ the product $\alpha\alpha^{\prime}$ is given by
\[
\alpha\alpha^{\prime}=\sum_{t=t_{0}+t_{0}}^{\infty}\left[\sum_{j=t_{0}}^{t-t_{0}^{o}}\alpha_{t}\alpha_{t-j}^{\prime}\right]z^{-t}
\]
and the sum $\alpha+\alpha^{\prime}$ is given by 
\[
\alpha+\alpha^{\prime}=\sum_{{t=\min^{\prime}\left(t_{0},t_{0}^{\prime}\right)\atop }}^{\infty}\left(\alpha_{i}+\alpha_{i}^{\prime}\right)z^{-t}
\]
With $\Lambda K$ as the underlying field it then follows that, with
convolution as the scalar multiplication and with the usual coefficientwise
addition, the set $\Lambda S$ becomes a $\Lambda K$ -linear space.
When $S$ is a finite dimensional $K$ -linear space, say of dimension
$n$ then so is $\Lambda S$ as a $\Lambda K$ -linear space. It is
readily observed that; under the same operations of convolution as
multiplication and coefficientwise addition, the field $\Lambda K$
contains (as subobjects) also (i) the ring $K[z]$, or in our notation
$\Omega^{+}K$, of polynomials in $z;$ (ii) the ring $K\left[\left[z^{-1}\right]\right]$,
or in our notation $\Omega^{-}K$, of formal power series in $z^{-1}$;
and finally, (iii) the field $K$ itself. It, thus, follows immediately
that the set $\Lambda S$ is not only a $\Lambda K$ -linear space
but is simultaneously also an $\Omega^{+}K$ -module, an $\Omega^{-}K$
-module and a $K$ -linear space As we shall see, these facts turn
out to be of central importance in the theory.

Now, we let $\mathbb{Z}$ denote the integers and for an element $s\in\Lambda S,$
given by $(2.1),$ we define the order of $s$ by
\[
(2.2)~~~~~~\ord s:=\left\{ \begin{array}{ll}
\min\left\{ t\in\mathbb{Z}|s_{t}\neq0\right\}  & \text{ if }s\neq0\\
\infty & \text{ if }s=0
\end{array}\right.
\]
If $s\neq0$ and $t_{0}=$ $\ord$ $s,$ we call the coefficient $s_{t_{0}}$
the leading coefficient of $s$. 

Let $U$ and $Y$ be $K$ -linear spaces. We shall call $U$ the input
value space and $Y$ the output value space of an underlying linear
system $\Sigma$. The $\Lambda K$ -linear spaces $\Lambda U$ and
$\Lambda Y$ are then called the extended input space and extended
output space, respectively Elements $u=\Sigma u_{i}z^{-t}\in\Lambda U$
and $y=\Sigma y_{i}z^{-t}\in\Lambda Y,$ called, respectively, (extended)
inputs and (extended) outputs, are identified with time sequences
$\left\{ u_{t}\right\} $ and $\left\{ y_{t}\right\} $ (with $t$
being identified as time marker).

Let $\bar{f}:\Lambda U\rightarrow\Lambda Y$ be a $K$ -linear map.
We say that $\bar{f}$ is time invariant if 
\[
\bar{f}(z\cdot u)=z\cdot\bar{f}(u)
\]
for all $u\in\Lambda U$, so that $\bar{f}$ is time invariant whenever
it is a $\Lambda K$ -linear map (Wyman $[1972]..$ Next, for a $\Lambda K$-linear
map $\bar{f}:\Lambda U\rightarrow\Lambda Y$ we define the order of
$\bar{f}$ by 
\[
(2.3)~~~~~\ord\bar{f}:=\inf\{\operatorname{ord}\bar{f}(u)-\operatorname{ord}u|0\neq u\in\Lambda U\}
\]
If the map $\bar{f}$ is the zero map then $\ord$ $\bar{f}:=\infty;$
otherwise $\ord$ $\bar{f}<\infty.$ While it is possible that $\ord$
$\bar{f}=-\infty$ we shall not concern ourselves here with this case
and confine our attention to maps of finite order. This is clearly
always the case when $U$ (and hence also $\Lambda U$ ) is finite
dimensional.

A $\Lambda K$ -linear map $\bar{f}:\Lambda U\rightarrow\Lambda Y$
is called causal if $\ord$ $\bar{f}\geqq0$ and strictly causal if
$\ord$ $\bar{f}>0.$ The map $\bar{f}$ is called order consistent
if for each $0\neq u\in\Lambda U$ 
\[
\operatorname{ord}\bar{f}(u)-\operatorname{ord}u=\operatorname{ord}\bar{f}
\]
Clearly, an invertible $\Lambda K$ -linear map $\bar{l}:\Lambda S\rightarrow\Lambda S$
is order consistent if and only if $\ord$ $\bar{l}^{-1}=-$ $\ord$
$\bar{l}$. A $\Lambda K$ -linear map $\bar{f}$ is said to be order
preserving (or instantaneous) if it is order consistent and $\ord$
$\bar{f}=0.$ An invertible order preserving (and hence causal) $\Lambda K$
-linear map $\bar{l}:\Lambda S\rightarrow\Lambda S$ is called $a$
bicausal isomorphism (or simply bicausal) since its inverse is then
also causal. Finally, we call $\bar{f}$ nonlatent if it is order
consistent and $\operatorname{ord}\bar{f}=1$.

We now introduce the following (see also Hautus and Heymann {[}1978{]}).
\begin{defn*}
2.4. A map $\bar{f}:\Lambda U\rightarrow\Lambda Y$ is called an extended
linear input-output map (or extended linear i/o map) if it is strictly
causal (i.e., $\ord$ $\bar{f}>0$ ) and $\Lambda K$ -linear.
\end{defn*}
Let $L$ denote the $K$ -linear space of $K$ -linear maps $U\rightarrow Y$
and let $\Lambda L$ denote the $\Lambda K$ -linear space of all
$L$ -Laurent series. We identify this space with the space of $\Lambda K$
-linear maps $\Lambda U\rightarrow\Lambda Y$ of finite order as follows.
We define the $K$ -linear maps 
\[
(2.5)~~~~~\begin{aligned}\bar{\imath}_{u} & :U\rightarrow\Lambda U:u\mapsto u\quad\text{(canonical injection)}\\
\bar{p}_{k} & :\Lambda Y\rightarrow Y:\Sigma y_{i}z^{-t}\mapsto y_{k}
\end{aligned}
\]
 and with every $\Lambda K$ -linear map $\bar{f}:\Lambda U\rightarrow\Lambda Y$
we associate the Laurent series 
\[
(2.6)~~~~~~Z_{\bar{f}}\left(z^{-1}\right):=\Sigma A_{t}z^{-t}
\]
 where, for each $k\in\mathbb{Z}$ 
\[
(2.7)~~~~~A_{k}:=A_{k}(\bar{f}):=\bar{p}_{k}\cdot\bar{f}\cdot\bar{\imath}_{u}
\]
The Laurent series (2.6) is called the impulse response or the transfer
function of $\bar{f}$. If $u=\Sigma u_{t}z^{-t}\in\Lambda U$ is
any element, then the action of $\bar{f}$ on $u$ is given by 
\[
(2.8)~~~~~\bar{f}\cdot u=\left(\Sigma A_{t}(\bar{f})z^{-t}\right)\cdot\left(\Sigma u_{t}z^{-t}\right)=\sum_{t}\sum_{k}\left(A_{k}(\bar{f})u_{t-k}\right)z^{-t}
\]
It is thus immediately seen that 
\[
(2.9)~~~~~\operatorname{ord}\bar{f}=\min\left\{ k|A_{k}(\bar{f})\neq0\right\} ,
\]
whence we have the following characterization of causality in terms
of the transfer function: The map $\bar{f}$ is causal if and only
if $A_{k}(\bar{f})=0$ for $k<0$ and strictly causal if and only
if $A_{k}(\bar{f})=0$ for $k\leqq0.$ We also have the following
easily verified proposition.
\begin{prop*}
2.10. Let $\bar{f}:\Lambda U\rightarrow\Lambda Y$ be a $\Lambda K$
-linear map of order $k_{0}(<\infty)$ and transfer function $Z_{\bar{f}}\left(z^{-1}\right)=\sum_{k=k_{0}}^{\infty}A_{k}z^{-k}.$
Then $\bar{f}$ is order consistent if and only if $A_{k_{0}}$ is
injective (i.e., $\ker$ $A_{k_{0}}=0$ ).
\end{prop*}
The following is an immediate corollary to Proposition 2.10.
\begin{cor*}
2.11. Let $\bar{l}:\Lambda S\rightarrow\Lambda S$ be a causal $\Lambda K$
-linear map with transfer function $\sum_{k=0}^{\infty}A_{k}(\bar{l})z^{-k}.$
Then $\bar{l}$ is a bicausal isomorphism if and only if $A_{0}(\bar{l})$
is invertible, in which case $A_{0}\left(\bar{l}^{-1}\right)=\left(A_{0}(\bar{l})\right)^{-1}$.
\end{cor*}
We associate with an extended linear i/o map $\bar{f}$ a restricted
linear i/ o map $\tilde{f}$ which is obtained as follows (see also
Hautus and Heymann {[}1978{]}). Inputs are restricted to the subset
$\Omega^{+}U\subset\Lambda U$, called the restricted input space,
and consist of all inputs that terminate at $t=0,$ i.e., elements
of the form $\Sigma_{t\leq0}u_{t}z^{-t}.$ Outputs are observed only
for $t\geqq1,$ that is, in the subset $z^{-1}\Omega^{-}Y$ which
is, of course, in bijective correspondence with the $\Omega^{+}K$
-quotient module $\Gamma^{+}Y:=\Lambda Y/\Omega^{+}Y$ which we call
the restricted output space. The restricted linear i/o map $\tilde{f}:\Omega^{+}U\rightarrow\Gamma^{+}Y$
associated with $\bar{f}$ is then defined by 
\[
\tilde{f}=\pi^{+}\cdot\bar{f}\cdot j^{+}
\]
where $j^{+}:\Omega^{+}U\rightarrow\Lambda U$ is the canonical injection
and $\pi^{+}:\Lambda Y\rightarrow\Gamma^{+}Y$ is the canonical projection.
Clearly, since $\pi^{+}$ and $j^{+}$ are $\Omega^{+}K$ -module
homomorphisms, so is also $\tilde{f}$ and we have the following:
\begin{defn*}
2.12. A map $\tilde{f}:\Omega^{+}U\rightarrow\Gamma^{+}Y$ is called
a restricted linear i/o map if it is an $\Omega^{+}K$ -module homomorphism.
\end{defn*}
Next, we define the linear output response (or output value) map $f:\Omega^{+}U\rightarrow Y$
associated with a given linear i/o map $\bar{f}$ (or $\tilde{f}$
) as follows: 
\[
(2.13)~~~~~f:\Omega^{+}U\rightarrow Y:u\mapsto f(u)=\bar{p}_{1}\cdot\bar{f}(u)=p_{1}\cdot\tilde{f}(u)
\]
 where (identifying $\left.\Gamma^{+}Y\text{ with }z^{-1}\Omega^{-}Y\right)$
\[
(2.14)~~~~~p_{1}:\Gamma^{+}Y\rightarrow Y:\sum_{t=1}^{\infty}y_{t}z^{-t}\mapsto y_{1}
\]
A linear i/o map $\bar{f}$ (or $\tilde{f}$ ) is called reachable
if the associated output value map $f$ is surjective.

If $f:\Omega^{+}U\rightarrow Y$ is any $K$ -linear map, it can be
regarded as an output value map of a linear system. In particular,
the restricted and extended linear i/o maps associated with $f$ are
then given by 
\[
(2.15)~~~~~\tilde{f}(u)=\sum_{t\geq0}f\left(z^{t}u\right)z^{-t-1},\quad u\in\Omega^{+}U
\]
and
\[
(2.16)~~~~~\bar{f}(u)=\sum_{t\in\mathbf{Z}}f\left(\mathscr{S}^{+}\left(z^{t}u\right)\right)z^{-t-1},\quad u\in\Lambda U
\]
where $\mathscr{S}^{+}:\Lambda U\rightarrow\Omega^{+}U:\Sigma u_{i}z^{-t}\mapsto\sum_{t\leq0}u_{i}z^{-t}$
is the truncation operator.

The relation between the maps $\bar{f},\tilde{f}$ and $f$ is summarized
by the commutative diagram, Fig. $2.1,$ in which $i$ denotes the
identity map. FIG. 2.1

\begin{figure}[h]
\begin{centering}
\includegraphics{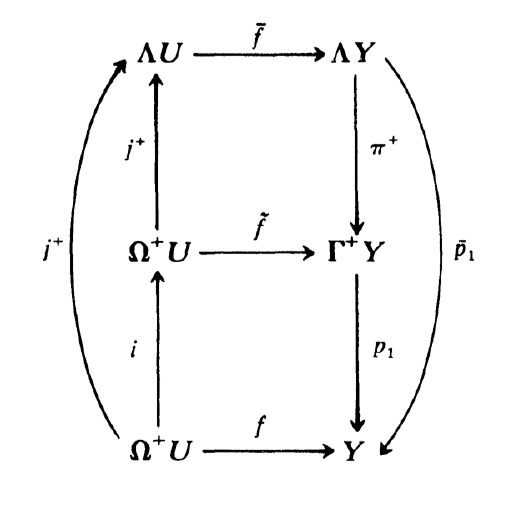}
\par\end{centering}
\caption{2.1}

\end{figure}

The output value map $f,$ which gives for each (restricted) input
the value of the output at time $t=1,$ is clearly a $K$ -linear
map. In some special cases, there exists an $\Omega^{+}K$ -module
structure on $Y$, compatible with its $K$ -vector space structure,
such that the output value map $f$ is not just $K$ -linear but is
also an $\Omega^{+}K$ -module homomorphism. When this is the case,
then for each $u\in\Omega^{+}U$ and for each positive integer $k,f\left(z^{k}u\right)=$
$z^{k}f(u),$ whence, by $(2.15),$ knowledge of the output value
at time $t=1$ implies knowledge of the whole ensuing output sequence.
This is therefore precisely the case when the system's output "qualifies"
as state, a fact which motivates the following definition (for greater
detail the reader is referred to Hautus and Heymann {[}1978{]}):
\begin{defn*}
$2.17.$ An extended linear i/o map $\bar{f}:\Lambda U\rightarrow\Lambda Y$
is called an extended linear input-state (or i/s) map if there exists
an $\Omega^{+}K$ -module structure on $Y$, compatible with its $K$
-linear structure, such that the output value map $f=\bar{p}_{1}\cdot\bar{f}\cdot j^{+}$
is an $\Omega^{+}K$ homomorphism. The associated restricted map $\tilde{f}$
is called $a$ restricted linear i/s map.
\end{defn*}
If $Y$ and $W$ are $K$ -linear spaces and $H:Y\rightarrow W$ is
a $K$ -linear map, then it induces in a natural way a $\Lambda K$
-linear map which we call static as follows:
\[
(2.18)~~~~~H:\Lambda Y\rightarrow\Lambda W:\Sigma y_{l}z^{-t}\mapsto\Sigma\left(Hy_{t}\right)z^{-t}
\]
In a similar way $H$ induces also static $\Omega^{+}K$ and $\Omega^{-}K$
-homomorphisms.

We shall need the following characterizations of linear i/s maps,
from Hautus and Heymann {[}1978{]}
\begin{thm*}
$2.19.$ If $\bar{f}:\Lambda U\rightarrow\Lambda Y$ is an extended
linear i/s map then 
\[
(2.20)~~~~~\operatorname{ker}f=\operatorname{ker}\tilde{f}.
\]
\end{thm*}
\begin{thm*}
2.21. Let $\bar{f}:\Lambda U\rightarrow\Lambda Y$ be a reachable
extended linear i/o map. Then the following are equivalent:
\begin{lyxlist}{MM}
\item [{(i)}] $\tilde{f}$ is an extended reachable linear i/s map.
\item [{(ii)}] Condition (2.20) holds
\item [{(iii)}] For every extended linear i/o map $\bar{g}:\Lambda U\rightarrow\Lambda W$
satisfying $\ker$ $\tilde{f}\subset$ $\ker$ $\tilde{g}$ (where
$\tilde{f}$ and $\tilde{g}$ are the corresponding restricted i/o
maps and where $W$ is a $K$ -linear space there exists a unique
static map $H:\Lambda Y\rightarrow\Lambda W$ such that $\bar{g}=H\cdot\bar{f}$.
\end{lyxlist}
\end{thm*}

\section*{3. Feedback and causal factorization-general considerations.}

We shall be concerned with the setup described by the block diagram
in Fig. 3.1.

\begin{figure}[h]
\begin{centering}
\includegraphics{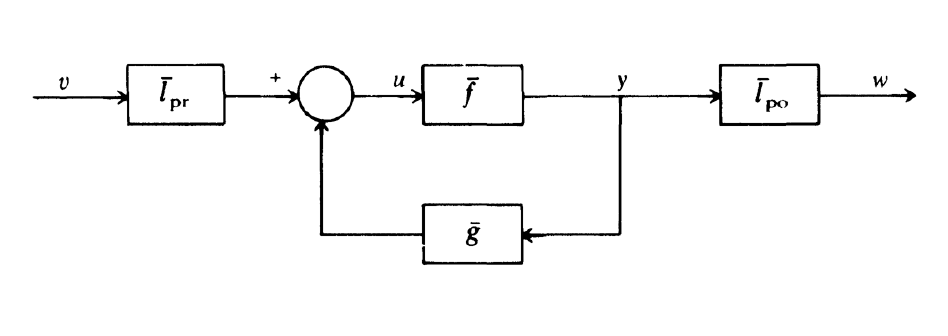}
\par\end{centering}
\caption{FIG. 3.1}

\end{figure}
Here $\bar{f}:\Lambda U\rightarrow\Lambda Y$ is an extended linear
i/o map, called the open loop system, $\bar{g}:\Lambda Y\rightarrow$
$\Lambda U$ is a causal $\Lambda K$ -linear map called the (output)
feedback compensator, $\bar{l}_{\mathrm{pr}}:\underline{\Lambda}U\rightarrow\Lambda U$
is a $\Lambda K$ -linear bicausal isomorphism called (bicausal) precompensator
and $\bar{l}_{\mathrm{po}}:\Lambda Y\rightarrow$ $\Lambda Y$ is
a $\Lambda K$ -linear bicausal isomorphism called (bicausal) postcompensator.
In case any of the maps $\bar{g},\bar{l}_{\mathrm{pr}}$ or $\bar{l}_{\mathrm{po}}$
is static we shall call it, respectively a static feedback, pre or
post compensator.

Now, since the map $\bar{g}$ is causal and $\bar{f}$ is strictly
causal, it readily follows that the composite maps $\bar{f}\cdot\bar{g}:\Lambda Y\rightarrow\Lambda Y$
and $\bar{g}\cdot\bar{f}:\Lambda U\rightarrow\Lambda U$ are both
strictly causal. Letting $I$ denote both of the corresponding identity
maps, we see that both of the maps $(I+\bar{g}\bar{f}):\Lambda U\rightarrow\Lambda U$
and $(I+\bar{f}\bar{g}):\Lambda Y\rightarrow\Lambda Y$ are bicausal
isomorphisms. It follows that the setup of Fig. 3.1 is "well-posed"
in the sense that there is a strictly causal $\Lambda K$ -linear
$\operatorname{map}\Lambda U\rightarrow\Lambda Y:v\mapsto w$ given
by either of the following composite maps: 
\[
(3.1)~~~~~v\mapsto w=\left[\bar{l}_{\mathrm{lo}}\cdot\bar{f}\cdot(I+\bar{g}\bar{f})^{-1}\cdot\bar{l}_{\mathrm{pr}}\right](v)
\]
\[
(3.2)~~~~~v\mapsto w=\left[\overline{l_{\mathrm{po}}}\cdot(I+\bar{f}\bar{g})^{-1}\cdot\bar{f}\cdot\bar{l}_{\mathrm{pr}}\right](v)
\]
Using again block diagrams, (3.1) and (3.2) can be described, respectively,
as in Fig. 3.2a and 3.2b.

In both descriptions, the dashed blocks represent bicausal mappings,
so that the compensator configuration of Fig. 3.1 can always be represented
equivalently

\begin{figure}[h]
\begin{centering}
\includegraphics{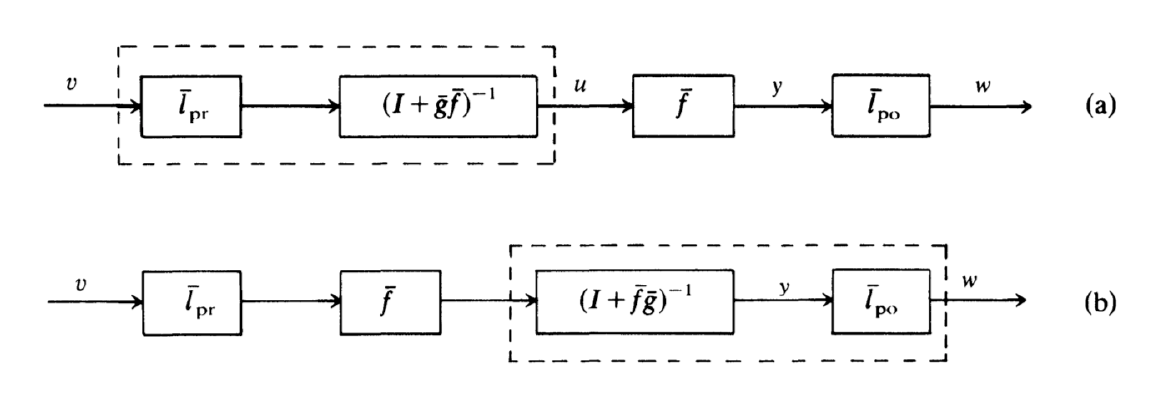}
\par\end{centering}
\caption{FIG. 3.2}

\end{figure}
by the original system preceded and followed by bicausal compensators,
with the feedback compensator represented, as one chooses, either
as a precompensator or a postcompensator.

Because of the obvious duality between the precompensator situation
and the postcompensator situation, there is no need to discuss both
of them in detail. since practical interest in postcompensators is
at best limited, we shall henceforth confine our attention to precompensation,
and discuss postcompensators only in connection with certain mathematical
questions.

For various reasons, not to be elaborated on here, feedback compensation
is preferred over external compensation whenever possible. Thus, one
is interested in the following problem.

\emph{Causal feedback problem $3.3.$} Let $\bar{f}:\Lambda U\rightarrow\Lambda Y$
be an extended linear i/o map.
\begin{lyxlist}{MM}
\item [{(a)}] Under what conditions can a given bicausal $\Lambda K$ -linear
isomorphism $\bar{l}:\Lambda\dot{U}\rightarrow$ $\Lambda U$ be represented
as feedback, i.e. under what conditions do there exist a static map
$L:\Lambda U\rightarrow\Lambda U$ and a causal $\Lambda K$ -linear
map $\bar{g}:\Lambda Y\rightarrow\Lambda Y$, such that $\bar{l}^{-1}=L+\bar{g}\bar{f}?$
\item [{(b)}] Under what conditions (on $\bar{f}$ ) can every bicausal
$\bar{l}$ be represented as feedback?
\end{lyxlist}
Let $\bar{l}:\Lambda U\rightarrow\Lambda U$ be a bicausal $\Lambda K$
-linear map, and let 
\[
Z_{I^{-1}\left(z^{-1}\right)}=\sum_{t=0}^{\infty}L_{t}z^{-t}
\]
denote the transfer function of $\bar{l}^{-1}$. We can then write
\[
Z_{I^{-1}}\left(z^{-1}\right)=L_{0}+\sum_{t=1}^{\infty}L_{t}z^{-t}=L_{0}+Z_{\bar{h}}\left(z^{-1}\right)
\]
where $L_{0}$ is a static $\Lambda K$ -linear map and $Z_{\bar{h}}\left(z^{-1}\right)$
is the transfer function of a strictly causal map $\bar{h}:\Lambda U\rightarrow\Lambda U$
representing the strictly causal part of $\bar{l}^{-1}$. Hence we
can always decompose the map $\bar{l}^{-1}$ as 
\[
\bar{l}^{-1}=L+\bar{h}
\]
with $L$ static and $\bar{h}$ strictly causal. The causal feedback
problem 3.3 is therefore essentially equivalent to the following.

\emph{Causal factorization problem $3.4.$} Let $\bar{f}:\Lambda U\rightarrow\Lambda Y$
be a given strictly causal $\Lambda K$ -linear map.
\begin{lyxlist}{MM}
\item [{(a)}] Under what conditions can a strictly causal $\Lambda K$
-linear map $\bar{h}:\Lambda U\rightarrow\Lambda U$ be factored causally
over $\bar{f},$ i.e., when does there exist a causal map $\bar{g}:\Lambda Y\rightarrow\Lambda U$
such that $\bar{h}=\bar{g}\cdot\bar{f}?$
\item [{(b)}] Under what conditions can every strictly causal $\Lambda K$
-linear map $\bar{h}:\Lambda U\rightarrow\Lambda U$ be factored causally
over $\bar{f}$ ? 
\end{lyxlist}
It is readily noted that the strict causality of the maps $\bar{f}$
and $\bar{h}$ is inessential to the causal factorization problem,
and arises in problem 3.4 only because of the specific requirements
of the feedback problem. Indeed, if $\bar{h}$ factors causally over
$\bar{f}$, i.e., if there exists a causal $\bar{g}$ such that $\bar{h}=\bar{g}\cdot\bar{f},$
then for each integer $k$ we also have $z^{k}\bar{h}=z^{k}\bar{g}\bar{f}=$
$\bar{g}\cdot\left(z^{k}\bar{f}\right)$ so that $z^{h}\bar{h}$ factors
causally over $z^{k}\bar{f}$, and for sufficiently large positive
$k$ (unless $\bar{h}$ or $\bar{f}$ are zero) the maps $z^{k}\bar{h}$
and $z^{k}\bar{f}$ are not causal. Thus, the causal factorization
problem can be stated in the following less restrictive way:

Given two $\Lambda K$ -linear maps $\bar{f}:\Lambda S\rightarrow\Lambda Y$
and $\bar{h}:\Lambda S\rightarrow\Lambda W$ (where $S,Y$ and $W$
are $K\text{ -linear spaces }),$ when does there exist a causal $\Lambda K$
-linear map $\bar{g}:\Lambda Y\rightarrow\Lambda W$ such that the
following diagram in Fig. 3.3 commutes

\begin{figure}[h]
\begin{centering}
\includegraphics{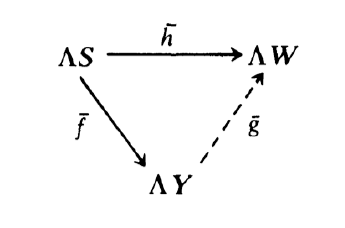}
\par\end{centering}
\caption{FIG. 3.3}

\end{figure}

FIG If the causality requirement of $\bar{g}$ is dropped, the factorization
problem is standard (see, e.g., Greub {[}1967{]}) and $\bar{h}$ factors
over $\bar{f}$ if and only if $\ker$ $\bar{f}\subset\mathrm{\ker}\bar{h}$.
Yet this condition does not say anything about the causality of $\bar{g}$.
To deal efficiently with the causality issue, we reintroduce the concept
of causality using an approach which is algebraically more tractable.

Let $\bar{f}:\Lambda U\rightarrow\Lambda Y$ be a $\Lambda K$ -linear
map. We can characterize causality of $\bar{f}$ as follows (compare
with our definitions of causality in $\S2$ ):
\begin{lyxlist}{00.00}
\item [{(3.5)}] The map $\bar{f}$ is causal if and only if $u\in\Omega^{-}U$
implies $\bar{f}(u)\in\Omega^{-}Y$.
\end{lyxlist}
Similarly, we have:
\begin{lyxlist}{00.00}
\item [{(3.6)}] The map $\bar{f}$ is strictly causal if and only if $u\in z\Omega^{-}U$
implies $\bar{f}(u)\in\Omega^{-}Y$.
\end{lyxlist}
Let us denote the $\Omega^{-}K$ -quotient module $\Lambda Y/\Omega^{-}Y$
by $\Gamma^{-}Y$, and let $\pi^{-}:\Lambda Y\rightarrow\Gamma^{-}Y$
denote the canonical projection. The following can then be easily
verified by the reader.
\begin{prop*}
3.7. Let $\bar{f}:\Lambda U\rightarrow\Lambda Y$ be a $\Lambda K$
-linear map.
\end{prop*}
\begin{lyxlist}{MM}
\item [{(a)}] The map $\bar{f}$ is causal if and only if $\Omega^{-}U\subset\operatorname{ker}\pi^{-}\bar{f}$.
\item [{(b)}] The map $\bar{f}$ is strictly causal if and only if $z\Omega^{-}U\subset$
$\ker$ $\pi^{-}\bar{f}$.
\item [{(c)}] The map $\bar{f}$ is order consistent if and only if, for
some integer $k,z^{k}\Omega^{-}U=\operatorname{ker}\pi^{-}\bar{f}$.
\item [{(d)}] The map $\bar{f}$ is instantaneous if and only if $\Omega^{-}U=\operatorname{ker}\pi^{-}\bar{f}$.
\item [{(e)}] The map $\bar{f}$ is nonlatent if and only if $z\Omega^{-1}U=\operatorname{ker}\pi^{-}\bar{f}$.
\end{lyxlist}
We shall use the characterizations of the above proposition extensively
in the following sections.

\section*{4. Proper independence and proper bases.}

Let $K$ be a field and let $S:=K^{m}$. For an element $0\neq s\in\Lambda S,$
denote by $\hat{s}$ the leading coefficient of $s.$ If $s=0$ we
shall say that $\hat{s}=0$.
\begin{defn*}
4.1. A set of vectors $s_{1},\cdots,s_{k}\in\Lambda S$ is called
properly independent if their leading coefficients $\hat{s}_{1},\cdots,\hat{s}_{k}\in S$
are $K$ -linearly independent.
\end{defn*}
Below we derive a variety of properties of properly independent sets,
of proper bases and of proper direct sum decompositions. Our objective
is to develop this theory here only to the extent required in the
sequel. Many further results have been omitted, and the reader can,
for example, easily verify that the converses of a number of our results
are also valid. A more extensive exposition of this and related topics
will be published elsewhere
\begin{lem*}
$4.2.$ If $s_{1},\cdots,s_{k}\in\Lambda S$ is a properly independent
set of vectors, then (i) it is $\Lambda K$-linearly independent,
and (ii) for every set of scalars $\alpha_{1},\cdots,\alpha_{k}\in\Lambda K$
the following holds 
\[
\operatorname{ord}\sum_{i=1}^{k}\alpha_{i}s_{i}=\min\left\{ \operatorname{ord}\alpha_{i}s_{i}|i=1,\cdots,k\right\} .
\]
\end{lem*}
Proof. We shall prove the lemma by showing that if either (i) or (ii)
fails to hold then the set $s_{1},\cdots,s_{k}$ is not properly independent.
If $\alpha_{1},\cdots,\alpha_{k}\in\Lambda K$ is any set of scalars
then, by definition, $\ord$ $\sum_{i=1}^{k}\alpha_{i}s_{i}\geqq r:=\min\left\{ \ord\alpha_{i}s_{i}|i=1,\cdots,k\right\} $.
If either (i) or (ii) fails to hold, there exist $\alpha_{1},\cdots,\alpha_{k}\in\Lambda K,$
not all zero, such that either $\sum_{i=1}^{k}\alpha_{i}s_{i}=0$
or $\ord$ $\sum_{i=1}^{k}\alpha_{i}s_{i}>r.$ For each $i=1,\cdots,k$
define
\[
\varepsilon_{i}:=\left\{ \begin{array}{ll}
1 & \text{ if }\ord\alpha_{i}s_{i}=r,\\
0 & \text{ if }\ord\alpha_{i}s_{i}>r,
\end{array}\right.
\]
and consider the terms of order $r$ in $\Sigma\alpha_{i}s_{i}$.
This yields $\sum_{i=1}^{k}\varepsilon_{i}\hat{\alpha}_{i}\hat{s}_{i}=0,$
implying that $\hat{s}_{1},\cdots,\hat{s}_{k}$ are $K$ -linearly
dependent since not all the $\varepsilon_{i}\hat{\alpha}_{i}$ are
zero. Hence $s_{1}\cdots,s_{k}$ are not properly independent, completing
the proof. $\square$

The condition of Lemma $4.2(\text{ ii })$ has been called the "predictable
degree property in Forney $[1975],$ in the (analogous) setting of
"minimal polynomial bases" for rational vector spaces. We shall
adopt this terminology and call the property of Lemma $4.2(\mathrm{ii})$
the \emph{predictable order property}.
\begin{defn*}
4.3. Let $\mathscr{R}\subset\Lambda S$ be a $\Lambda K$ -linear
subspace. A basis $\left\{ s_{1},\cdots,s_{k}\right\} $ of $\mathscr{R}$
is called proper if the vectors $s_{1},\cdots,s_{k}$ are properly
independent. The basis is called normalized if for each $i=1,\cdots,k,$
$\ord$ $s_{i}=0$.
\end{defn*}
To avoid possible confusion in the ensuing discussion where we shall
deal with both $K$ -linear and $\Lambda K$ -linear spaces, we shall
use subscripts to emphasize the field. Thus, for example, $\operatorname{span}_{\Lambda K}\left\{ s_{1},\cdots,s_{k}\right\} $
denotes the $\Lambda K$ -linear subspace spanned by $s_{1},\cdots,s_{k}\in$
$\Lambda S,$ whereas $\operatorname{span}_{K}\left\{ \hat{s}_{1},\cdots,\hat{s}_{k}\right\} $
denotes the $K$ -linear subspace spanned by $\hat{s}_{1},\cdots,\hat{s}_{k}\in$
$S.$ Similarly, $\operatorname{dim}_{\Lambda K}\mathscr{R}$ denotes
the dimension of a subspace $\mathscr{R}\subset\Lambda S$ as a $\Lambda\mathrm{K}$
-linear space (to distinguish from $K$ -linear). We next have the
following theorem.
\begin{thm*}
4.4. Every nonzero $\Lambda K$-linear subspace $\mathscr{R}\subset\Lambda S$
has a proper basis. Moreover, every properly independent subset of
$\mathscr{R}$ can be extended to a proper basis.
\end{thm*}
Proof. Let $0\neq s_{1}\in\mathscr{R}$ be any vector. Then $s_{1}$
is properly independent. We shall complete the proof by showing that
if $s_{1},\cdots,s_{k}\in\mathscr{R}$ are a properly independent
set and if $\mathscr{R}_{k}:=\operatorname{span}_{\Lambda K}\left\{ s_{1},\cdots,s_{k}\right\} $
is a proper subspace of $\mathscr{P},$ we can find a vector $s_{k+1}\in\mathscr{R}$
such that the set $\left\{ s_{1},\cdots,s_{k},s_{k+1}\right\} $ is
also properly independent. The proof is by contradiction. Assume that
$\mathscr{R}_{k}\subset\mathscr{R}$ is a proper subspace, let $s_{k+1}^{\circ}\in\mathscr{R}$
be such that the set $\left\{ s_{1},\cdots,s_{k},s_{k+1}^{\circ}\right\} $
is $\Lambda K$ -linearly independent and, without loss of generality,
assume that this set is also normalized. Let $\mathscr{P}_{k+1}:=\operatorname{span}_{\Lambda K}\left\{ s_{1},\cdots,s_{k},s_{k+1}^{\circ}\right\} $
and suppose that there is no vector $s\in\mathscr{R}_{k+1}$ such
that the set $\left\{ s_{1},\cdots,s_{k},s\right\} $ is properly
independent. This means that for each $s\in\mathscr{R}_{k+1},\hat{s}\in\widehat{\mathscr{R}}_{k}:=\operatorname{span}_{K}\left\{ \hat{s}_{1},\cdots,\hat{s}_{k}\right\} ,$
contradicting, as we shall see, the $\Lambda K$ -linear independence
of $s_{1},\cdots,s_{k},s_{k+1}^{\circ}$. Indeed, we observe that
there are scalars $\alpha_{1}^{\circ},\cdots,\alpha_{k}^{\circ}\in K$
such that $\hat{s}_{k+1}^{\circ}=\sum_{i=1}^{k}\alpha_{i}^{\circ}\hat{s}_{i}$,
Let $n_{0}:=0$ and set $s_{k+1}^{1}:=s_{k+1}^{\circ}-$ $\sum_{i=1}^{k}\alpha_{i}^{\circ}z^{-n_{S}}s_{i},$
so that $\ord$ $s_{k+1}^{1}>$ $\ord$ $s_{k+1}^{\circ}.$ We now
form a sequence of vectors $\left\{ s_{k+1}^{t}\right\} $ $t=0,1,2,\cdots,$
with $s_{k+1}^{t}\in\mathscr{R}_{k+1},$ such that $\ord$ $s_{k+1}^{t+1}>$
$\ord$ $s_{k+1}^{t}$ for all $t=0,1,2,\cdots$ as follows: For each
$t$, set $n_{t}=\operatorname{ord}s_{k+1}^{t}$ and let $s_{k+1}^{r+1}:=s_{k+1}^{t}-\sum_{i=1}^{k}\alpha_{i}^{t}z^{-n}s_{i},$
where the scalars $\alpha_{1}^{t},\cdots,\alpha_{k}^{t}\in K$ satisfy
the condition that $\hat{s}_{k+1}^{t}=\sum_{i=1}^{k}\alpha_{i}^{t}\hat{s}_{i}$.
Upon defining $\alpha_{i}:=\sum_{t=0}^{\infty}\alpha_{i}^{t}z^{-n_{i}}\in\Lambda K,\quad i=1,\cdots,k,$
it is readily verified that $s_{k+1}^{\circ}-\sum_{i=1}^{k}\alpha_{i}s_{i}=0$
whence $s_{k+1}^{\circ}\in\mathscr{R}_{k},$ a contradiction. $\square$ 
\begin{cor*}
4.5. Let $\mathscr{R}\subset\Lambda S$ be a $\Lambda K$ -linear
subspace. Then $\operatorname{dim}_{\Lambda K}\mathscr{R}=\operatorname{dim}_{K}\widehat{\mathscr{R}}$
where $\hat{\mathscr{R}}:=\operatorname{span}_{K}\{\hat{s}|s\in\mathscr{R}\}$.
\end{cor*}
Let $\mathscr{R}\subset\Lambda S$ be a $\Lambda K$ -linear subspace.
If $\mathscr{R}=\mathscr{R}_{1}\oplus\mathscr{R}_{2}$ is a direct
sum decomposition of $\mathscr{R}$ into $\Lambda K$ -linear subspaces
$\mathscr{R}_{1}$ and $\mathscr{P}_{2}$, then, in general, $\hat{\mathscr{R}}_{1}\cap\hat{\mathscr{R}}_{2}\neq0$
so that $\hat{\mathscr{R}}\neq\hat{\mathscr{R}}_{1}+\hat{\mathscr{R}}_{2}.$
This leads us to the following
\begin{defn*}
4.6. A direct sum decomposition $\mathscr{R}=\mathscr{R}_{1}\oplus\mathscr{R}_{2}$
of a $\Lambda K$ -linear subspace $\mathscr{R}\subset\Lambda S$
into $\Lambda K$ -linear subspaces $\mathscr{R}_{1}$ and $\mathscr{R}_{2}$
is called proper if $\hat{\mathscr{R}}_{1}\cap\hat{\mathscr{R}}_{2}=0$
The subspace $\mathscr{P}_{2}$ is then called a \emph{proper direct
summand} of $\mathscr{P}_{1}$,
\end{defn*}
With the aid of Corollary 4.5 it is readily seen that a direct sum
decomposition is proper if and only if $\hat{\mathscr{R}}=\hat{\mathscr{R}}_{1}+\hat{\mathscr{R}}_{2}.$
Thus, $\mathscr{R}=\mathscr{R}_{1}\oplus\mathscr{P}_{2}$ is a proper
decomposition if and only if there are proper bases $s_{11},\cdots,s_{1k_{1}}$
of $\mathscr{R}_{1}$ and $s_{21},\cdots s_{2k_{2}}$ of $\mathscr{R}_{2}$
such that the set $s_{11},\cdots,s_{1k_{1}},s_{21},\cdots,s_{2k_{2}}$
is a proper basis of $\mathscr{R}.$ We then have the following further
corollary to Theorem 4.4.
\begin{cor*}
4.7. Let $\mathscr{R}\subset\Lambda S$ be a $\Lambda K$ -linear
subspace. Then every $\Lambda K$ -linear subspace $\mathscr{R}_{1}\subset\mathscr{R}$
has a proper direct summand in $\mathscr{R}$.
\end{cor*}
Finally, we also have the following variant of the predictable order
property.
\begin{cor*}
4.8. Let $\mathscr{R}=\mathscr{R}_{1}\oplus\mathscr{R}_{2}$ be a
proper direct sum decomposition of a MK-linear subspace $\mathscr{R}\subset\Lambda$
S. Let $s=s_{1}+s_{2}$ be the representation of any vector $s\in\mathscr{R},$
with $s_{i}\in\mathscr{R}_{i},i=1,2.$ Then $\ord s=\min\left\{ \ord s_{1},\ord s_{2}\right\} $.
\end{cor*}
Proof. By definition, $\ord$ $s\geqq\min\left\{ \ord s_{1},\ord s_{2}\right\} .$
If the above inequality is strict there exist scalars $\alpha_{1},\alpha_{2}\in K,$
not both zero, such that $\alpha_{1}\hat{s}_{1}+\alpha_{2}\hat{s}_{2}=0$
contradicting the fact that $\hat{\mathscr{R}}_{1}\cap\hat{\mathscr{R}}_{2}=0$.~~~$\square$

\section*{5. Causal factorization.}

We turn now to the causal factorization problem (3.4). As we mentioned
earlier, there is no essential need, in characterizing causal factorizability,
to assume strict causality, or even causality, of the maps under consideration.
We shall therefore begin with the general case and turn to specific
consideration of i/o maps later on. We shall assume that the spaces
$U$ and $Y$ are finite dimensional, in particular that $U=K^{m}$
and $Y=K^{p}.$ For convenience of notation, we shall temporarily
use the notation $\Lambda U$ and $\Lambda Y$ also in connection
with $\Lambda K$ -linear maps $\bar{f}:\Lambda U\rightarrow\Lambda Y$
that are not necessarily i/o maps (i.e., are not necessarily strictly
causal).

Let $\bar{f}:\Lambda U\rightarrow\Lambda Y$ be a $\Lambda K$ -linear
map and let $\pi^{-}:\Lambda Y\rightarrow\Gamma^{-}Y:=\Lambda Y/\Omega^{-}Y$
be the canonical projection. since $\Omega^{-}Y$ is an $\Omega^{-}K$
-module, so is the quotient $\Lambda Y/\Omega^{-}Y$. Thus the map
$\pi^{-}$ is an $\Omega^{-}K$ -homomorphism and so is also the composite
$\pi^{-}\bar{f}$. We have
\begin{lem*}
5.1. Let $\bar{f}:\Lambda U\rightarrow\Lambda Y$ be a $\Lambda K$
-linear map and let $\pi^{-}:\Lambda Y\rightarrow\Gamma^{-}Y$ be
the canonical projection. If $\mathscr{R}\subset$ $\ker$ $\pi^{-}\bar{f}$
is a $\Lambda K$ -linear subspace, then $\mathscr{R}\subset\mathrm{\ker}\bar{f}$.
\end{lem*}
Proof. Assume $u\in\mathscr{R}\subset\mathrm{\ker}\pi^{-}\bar{f},$
where $\mathscr{R}$ is a $\Lambda K$ -linear subspace. Then $\alpha u\in$
$\ker$ $\pi^{-}\bar{f}_{-}$ for all $\alpha\in\Lambda K.$ Thus
$\bar{f}(\alpha u)=\alpha\bar{f}(u)\in\Omega^{-}Y$ for all $\alpha\in\Lambda K,$
whence $\bar{f}(u)=0$ and $u\in\operatorname{ker}\bar{f}$ as claimed.
$\square$

Next we have the following central theorem.
\begin{thm*}
5.2. Let $\bar{f}:\Lambda U\rightarrow\Lambda Y$ and $\bar{h}:\Lambda U\rightarrow\Lambda W$
be $\Lambda K$ -linear maps, where $U$ Y and W are finite dimensional
K-linear spaces. There exists a causal $\Lambda K$ -linear map $\bar{g}:\Lambda Y\rightarrow\Lambda W$
such that $\bar{h}=\bar{g}\cdot\bar{f}$ if and only if $\ker$ $\pi^{-}\bar{f}\subset\operatorname{ker}\pi^{-}\bar{h}$.
\end{thm*}
Proof. Suppose $\bar{h}=\bar{g}\cdot\bar{f}$ with $\bar{g}$ causal.
Let $u\in\operatorname{ker}\pi^{-}\bar{f}.$ Then $\bar{f}(u)\in\Omega^{-}Y,$
and by causality of $\bar{g}$ (see Proposition 3.7(a)) $\Omega^{-}Y\subset$
$\ker$ $\pi^{-}\bar{g}$. It follows that $\bar{f}(u)\in\operatorname{ker}\pi^{-}\bar{g}$
whence $u\in\operatorname{ker}\pi^{-}\bar{g}\cdot\bar{f}=\operatorname{ker}\pi^{-}\bar{h}.$
Conversely, assume that $\ker$ $\pi^{-}\bar{f}\subset\mathrm{\ker}\pi^{-}\bar{h}.$
By Lemma 5.1 this implies that $\ker$ $\bar{f}\subset$ $\ker$ $\bar{h}$
whence by a standard theorem of linear algebra (see, e.g., Greub {[}1967{]})
a $\Lambda K$ -linear map $\bar{g}:\Lambda Y\rightarrow\Lambda W$
such that $\bar{h}=\bar{g}\cdot\bar{f}$ exists. It remains to be
shown that the map $\bar{g}$ can be selected to be causal. To this
end write $\Lambda Y=\operatorname{Im}\bar{f}\oplus\mathscr{R},$
where $\operatorname{Im}\bar{f}$ is the image of $\bar{f}$ and $\mathscr{P}$
is any proper direct summand (see Corollary 4.7 ). Let $\bar{g}_{0}:\Lambda Y\rightarrow\Lambda W$
be any $\Lambda K$ -linear map that satisfies the condition that
$\bar{h}=\bar{g}_{0}\cdot\bar{f}$ and let $\bar{g}_{1}:\operatorname{Im}\bar{f}\rightarrow\Lambda W$
be the restriction of $\bar{g}_{0}$ to the image of $\bar{f}$. Let
$p:\Lambda Y\rightarrow\operatorname{Im}\bar{f}$ denote the projection
onto $\operatorname{Im}\bar{f}$ along $\mathscr{P};$ that is, if
$y=y_{1}+y_{2}\in\Lambda Y$ is the decomposition of $y$ into its
components $y_{1}\in\operatorname{Im}\bar{f}$ and $y_{2}\in\mathscr{R},$
then $py=y_{1}.$ Clearly, $p$ is $\Lambda K$ -linear, and we shall
see that the map $\bar{g}=\bar{g}_{1}\cdot p$ satisfies the conditions
of the theorem. First observe that for $u\in\Lambda U$ 
\[
\bar{g}\cdot\bar{f}(u)=\bar{g}_{1}\cdot p\bar{f}(u)=\bar{g}_{0}\bar{f}(u)=\bar{h}(u)
\]
so that $\bar{g}\cdot\bar{f}=\bar{h}.$ To see that $\bar{g}$ is
causal, let $y=y_{1}+y_{2}\in\Omega^{-}Y,$ where $y_{1}\in\operatorname{Im}\bar{f}$
and $y_{2}\in\mathscr{R}.$ By Proposition $3.7(\mathrm{a}),$ the
proof will be complete if we show that $y\in\mathrm{\ker}\pi^{-}\bar{g}$
Indeed, Corollary 4.8 implies that both $y_{1}$ and $y_{2}$ are
in $\Omega^{-}Y$ so that $\bar{g}\cdot y=\bar{g}_{1}\cdot py=$ $\bar{g}_{1}\cdot y_{1}=\bar{g}_{0}\cdot\bar{f}(u)$
for some $u\in\operatorname{ker}\pi^{-}\bar{f}.$ But by hypothesis
$\ker$ $\pi^{-}\bar{f}\subset$ $\ker$ $\pi^{-}\bar{h},$ whence
$\bar{g}\cdot y=\bar{g}_{0}\cdot\bar{f}(u)=\bar{h}(u)\in\Omega^{-}W$
so that $y\in\operatorname{ker}\pi^{-}\bar{g}$ as claimed. $\square$

Theorem 5.2 clarifies the significance of the $\Omega^{-}K$ -module
$\ker$ $\pi^{-}\bar{f}$ in connection with the causal factorization
problem (and consequently also with feedback). We call this module
the \emph{latency module} or \emph{latency kernel} of $\bar{f}$.

COROLLARY 5.3. Let $\bar{f}:\Lambda U\rightarrow\Lambda Y$ be a $\Lambda K$
-linear map of finite order. Then $\bar{f}$ is order consistent if
and only if for every $\Lambda K$ -linear map $\bar{h}:\Lambda U\rightarrow\Lambda W$
which satisfies $\ord$ $\bar{h}\geqq$ $\ord$ $\bar{f}$ there exists
a causal $\Lambda K$ -linear map $\bar{g}:\Lambda Y\rightarrow\Lambda W$
such that $\bar{h}=\bar{g}\cdot\bar{f}$.

Proof. Recall that a map $\bar{f}$ is order consistent if $\ord$
$\bar{f}(u)-$ $\ord$ $u=$ $\ord$ $\bar{f}$ for each $0\neq u\in\Lambda U.$
Suppose $\bar{f}$ is order consistent and $\ord$ $\bar{h}\geqq\operatorname{ord}\bar{f}.$
Let $0\neq u\in\operatorname{ker}\pi^{-}\bar{f}.$ Then $\bar{f}(u)\in\Omega^{-}Y$
and $\ord$ $\bar{f}(u)\geqq0.$ Now $\ord$ $\bar{h}(u)-$ $\ord$
$u\geqq\operatorname{ord}\bar{h}\geqq\operatorname{ord}\bar{f}=\operatorname{ord}\bar{f}(u)-\operatorname{ord}u$
whence $\ord$ $\bar{h}(u)\geqq\operatorname{ord}\bar{f}(u)\geqq0,$
so that $u\in\operatorname{ker}\pi^{-}\bar{h},$ implying that $\operatorname{ker}\pi^{-}\bar{f}\subset\operatorname{ker}\pi^{-}\bar{h}$
By Theorem 5.2 the existence of a causal $\bar{g}$ such that $\bar{h}=\bar{g}\cdot\bar{f}$
is thus assured. Conversely, suppose $\bar{f}$ is not order consistent
and that $\bar{h}$ is an order consistent map satisfying $\ord$
$\bar{h}=\operatorname{ord}\bar{f}.$ Then there exists $0\neq u\in\Lambda U$
such that $\ord$ $\bar{f}(u)>\operatorname{ord}\bar{f}+$ $\ord$
$u=\operatorname{ord}\bar{h}+\operatorname{ord}u=\operatorname{ord}\bar{h}(u).$
If $k:=\operatorname{ord}\bar{f}(u),$ then $0=\operatorname{ord}\bar{f}\left(z^{k}u\right)>\operatorname{ord}\bar{h}\left(z^{k}u\right)$
so that $z^{k}u\in\operatorname{ker}\pi^{-}\bar{f}$ but $z^{k}u\notin$
$\ker$ $\pi^{-}\bar{h}.$ Hence $\ker$ $\pi^{-}\bar{f}\notin$ $\ker$
$\pi^{-}\bar{h}$ and by Theorem 5.2 there does not exist a causal
$\bar{g}$ such that $\bar{h}=\bar{g}\cdot\bar{f},$ completing the
proof. $\square$

The following corollary which is an immediate consequence of Corollary
5.3 is of central interest in our study of causal factorization since
it deals with linear i/o maps and gives us an important characterization
of nonlatency.
\begin{cor*}
5.4. Let $\bar{f}:\Lambda U\rightarrow\Lambda Y$ be an extended linear
i/o map. Then $\bar{f}$ is nonlatent if and only if for every strictly
causal $\Lambda K$ -linearmap $\bar{h}:\Lambda U\rightarrow\Lambda W$
there exists $a$ causal $\Lambda K$ -linear map $\bar{g}:\Lambda Y\rightarrow\Lambda W$
such that $\bar{h}=\bar{g}\cdot\bar{f}$.
\end{cor*}
Let $\bar{f}:\Lambda U\rightarrow\Lambda Y$ be an extended linear
i/o map and let $\bar{l}:\Lambda U\rightarrow\Lambda U$ be a bicausal
isomorphism, i.e., a bicausal precompensator for $\vec{f}$. Let $\bar{h}$
be the strictly causal part of $\bar{l}^{-1},$ i.e., $\bar{l}^{-1}=L+\bar{h}$
where $L$ is static. As we have seen in $\S3,\bar{l}$ can be realized
as feedback around $\bar{f}$ if $\bar{h}$ factors causally over
$\bar{f}$. Theorem 5.2 tells us essentially that the only barrier
to realizing a bicausal precompensator as feedback is the relative
latency of $\bar{f}$ and $\bar{h}.$ Corollary 5.4 characterizes
the class of i/o maps over which every bicausal precompensator can
be realized as feedback. These i/o maps are, as we have seen, the
nonlatent maps (a fact which motivated our choice of terminology).
Now, a very special and important class of nonlatent maps is that
of injective i/s maps. This fact is proved in the following theorem.
\begin{thm*}
5.5. Let $\bar{f}:\Lambda U\rightarrow\Lambda Y$ be an injective
linear i/s map. Then $\bar{f}$ is nonlatent. 
\end{thm*}
Proof. By strict causality of $\bar{f}$ we have that $z\Omega^{-}U\subset$
$\ker$ $\pi^{-}\bar{f},$ so that to prove nonlatency we need only
to show that $\ker$ $\pi^{-1}\bar{f}\subset z\Omega^{-}U$. Let $u\in$
$\ker$ $\pi^{-}\bar{f}$ so that $\bar{f}(u)\in\Omega^{-}Y.$ Write
$u=u^{+}+u^{-},$ where $u^{+}\in z^{2}\Omega^{+}U$ and $u^{-}\in z\Omega^{-}U.$
The proof will be completed by showing that $u^{+}=0$ so that $u\in z\Omega^{-}U$
as claimed. Note that $\bar{f}\left(u^{-}\right)\in\Omega^{-}Y$ by
the strict causality of $\bar{f}$ so that, in view of the fact that
$\bar{f}(u)=\bar{f}\left(u^{+}\right)+\bar{f}\left(u^{-}\right),$
it follows that $\bar{f}\left(u^{+}\right)\in\Omega^{-}Y.$ By (2.16)
we have
\[
\bar{f}\left(u^{+}\right)=\sum_{t\in\mathbf{Z}}f\left(\mathscr{S}^{+}\left(z^{i}u^{+}\right)\right)z^{-t-1}\in\Omega^{-}Y,
\]
so that, in particular, $f\left(\mathscr{S}^{+}\left(z^{-2}u^{+}\right)\right)=0.$
But $z^{-2}u^{+}\in\Omega^{+}U,$ whence $f\left(\mathscr{S}^{+}\left(z^{-2}u^{+}\right)\right)=$
$f\left(z^{-2}u^{+}\right)=0$ implying that $z^{-2}u^{+}\in\operatorname{ker}f=\operatorname{ker}\tilde{f}$
(the equality being a consequence of the i/s property (2.20) ). It
follows that $\bar{f}\left(z^{-2}u^{+}\right)\in\Omega^{+}Y$, or
alternatively, that $\bar{f}\left(u^{+}\right)\in$ $z^{2}\Omega^{+}Y.$
since $z^{2}\Omega^{+}Y\cap\Omega^{-}Y=0,$ we conclude that $\bar{f}\left(u^{+}\right)=0$
or that $u^{+}=0$ by the injectivity of $\bar{f}$. $\square$

While Theorem 5.5 deals only with injective i/s maps, it is important
to observe that this is not a serious restriction. Indeed, it is shown
in Proposition 5.6 below that in the special case of i/s maps (in
contrast to i/o maps in general), the kernel is "static"; i.e.,
if $\bar{f}$ is a noninjective i/s map, then $\operatorname{ker}\bar{f}=\Lambda U^{0}$
where $U^{0}\subset U$ is a subspace. This means that the whole degeneracy
lies in the input value space $U$ which has been chosen too large,
and by restricting the input value space to a proper summand of $U^{0}$
in $U$, the injectivity is restored.
\begin{prop*}
5.6. Let $\bar{f}:\Lambda U\rightarrow\Lambda Y$ be an extended linear
i/s map. Then there exists a subspace $U^{0}\subset U$ such that
$\operatorname{ker}\bar{f}=\Lambda U^{0}$.
\end{prop*}
Proof. Let $i_{u}:U\rightarrow\Omega^{+}U:u\mapsto u$ be the canonical
injection and define the subspace $U^{0}\subset U$ as $U^{0}:=$
$\ker$ $f\cdot i_{u},$ where $f$ is the output value map associated
with $\bar{f}$. since $\bar{f}$ is an i/s map we have $\ker$ $f\cdot i_{u}=\operatorname{ker}\tilde{f}\cdot i_{u}=\operatorname{ker}\bar{f}\cdot\bar{i}_{u}$
with the last equality holding by the strict causality of $\bar{f}$.
Thus $\bar{i}_{u}\left(U^{0}\right)\subset$ $\ker$ $\bar{f}$, and
since $\ker$ $\bar{f}$ is a $\Lambda K$ -linear space we conclude
that $\Lambda U^{0}\subset$ $\ker$ $\bar{f}$. To prove that $\ker$
$\bar{f}\subset\Lambda U^{0}$, it suffices to prove that if $0\neq u=\sum_{t=t_{0}}^{\infty}u_{t}z^{-t}\in\operatorname{ker}\bar{f}$
then $u_{t_{0}}\in U^{0}.$ By recursive application of the same argument
this will then imply that $u_{t}\in U^{0}$ for all $t\geqq t_{0}.$
Now by formula (2.16) we have $f\left(\mathscr{S}^{+}\left(z^{k}u\right)\right)=0$
for all $k\in\mathbb{Z},$ and since $\mathscr{S}^{+}\left(z^{t_{0}}u\right)=u_{t_{0}}$
the results follow. $\square$

The importance of Theorem 5.5 lies in the fact that it tells us that
bicausal precompensation is equivalent, in the sense of solvability,
to dynamic state feedback Let $\bar{f}:\Lambda U\rightarrow\Lambda Y$
be an extended linear i/o map. We write (see Hautus and Heymann {[}1978{]})
$\bar{f}=H\cdot\bar{f}_{s},$ where $H$ is a static output map and
$\bar{f}_{s}$ is a reachable i/s map. If $\bar{f}_{s}$ is injective
(which is always the case when $\ker$ $\bar{f}$ does not contain
a subspace of the form $\Lambda S,0\neq S\subset U$ ), then every
bicausal precompensator can be realized as feedback around $\bar{f}_{s}$.
That is, we can write every bicausal $\bar{l}:\Lambda U\rightarrow\Lambda U$
as $\bar{l}^{-1}=L+\bar{g}\bar{f}_{s},$ where $\bar{g}:\Lambda Y\rightarrow\Lambda U$
is a causal $\Lambda K$ -linear map and $L$ is static.

Before we proceed with our general investigation, it is worthwhile
to record one more consequence of Theorem 5.2.
\begin{cor*}
5.7. Let $\bar{f}_{1},\bar{f}_{2}:\Lambda U\rightarrow\Lambda Y$
be two extended linear i/o maps with $U$ and Y finite dimensional
K-linear spaces. There exists a bicausal $\Lambda K$ -linearmap $\bar{l}:\Lambda Y\rightarrow\Lambda Y$
such that $\bar{f}_{2}=\bar{l}\cdot\bar{f}_{1}$ if and only if $\ker$
$\pi^{-}\bar{f}_{1}=\operatorname{ker}\pi^{-}\bar{f}_{2}$.
\end{cor*}
Proof. First, observe that if a bicausal $\bar{l}$ exists then, by
Theorem $5.2,$ it follows immediately that $\ker$ $\pi^{-}\bar{f}_{1}=\operatorname{ker}\pi^{-}\bar{f}_{2}$.
Conversely, assume that $\ker$ $\pi^{-}\bar{f}_{1}=\operatorname{ker}\pi^{-}\bar{f}_{2}$
and write $\Lambda Y=\operatorname{Im}\bar{f}_{1}\oplus\mathscr{R}_{1}=\operatorname{Im}\bar{f}_{2}\oplus\mathscr{R}_{2}$
where $\mathscr{R}_{1}$ and $\mathscr{R}_{2}$ are proper direct
summands By Theorem 5,2 there exist causal maps $\bar{l}^{1},\bar{l}^{2}:\Lambda Y\rightarrow\Lambda Y$
such that $\bar{l}^{1}\bar{f}_{1}=\bar{f}_{2}$ and $\bar{l}^{2}\bar{f}_{2}=\bar{f}_{1}.$
Hence $\bar{l}^{2}\cdot\bar{l}^{1}\bar{f}_{1}=\bar{f}_{1},$ and letting
$\bar{l}_{1}:\operatorname{Im}\bar{f}_{1}\rightarrow\Lambda Y$ denote
the restriction of $\bar{l}^{1}$ to the image of $\bar{f}_{1},$
it is readily verified that $\bar{l}_{1}$ is order preserving. Now,
$\ker$ $\pi^{-}\bar{f}_{1}=$ $\ker$ $\pi^{-}\bar{f}_{2}$ implies
that $\operatorname{ker}\bar{f}_{1}=\operatorname{ker}\bar{f}_{2},$
whence $\operatorname{dim}\operatorname{Im}\bar{f}_{1}=\operatorname{dim}\operatorname{Im}\bar{f}_{2}$
and $\operatorname{dim}\mathscr{R}_{1}=$ $\operatorname{dim}\mathscr{R}_{2}.$
Let $\bar{l}_{2}:\mathscr{R}_{1}\rightarrow\Lambda Y$ be an order
preserving map satisfying $\operatorname{Im}\bar{l}_{2}=\mathscr{R}_{2}$
and let $p:\Lambda Y\rightarrow\operatorname{Im}\bar{f}_{1}$ denote
the projection along $\mathscr{R}_{1}.$ We claim that the map $\bar{l}:\Lambda Y\rightarrow\Lambda Y$
: $y\mapsto\bar{l}_{1}py+\bar{l}_{2}(I-p)y$ is a bicausal isomorphism
and that $\bar{l}\cdot\bar{f}_{1}=\bar{f}_{2}$. Indeed, to see the
latter property, note that for any $u\in\Lambda U$ we have 
\[
\bar{l}\bar{f}_{1}(u)=\bar{l}_{1}p\bar{f}_{1}(u)+\bar{l}_{2}(I-p)\bar{f}_{1}(u)=\bar{l}_{1}\cdot\bar{f}_{1}(u)=\bar{l}^{1}\bar{f}_{1}(u)=\bar{f}_{2}(u).
\]
To see the bicausality of $\bar{l}$ it suffices to show that it is
order preserving. Indeed, let $y=y_{1}+y_{2}\in\Lambda Y$ be any
element with $y_{1}\in\operatorname{Im}\bar{f}_{1}$ and $y_{2}\in\mathscr{R}_{1}$.
Then $\bar{l}y=\bar{l}_{1}y_{1}+\bar{l}_{2}y_{2}$ and using Corollary
4.8 together with the fact that $\operatorname{Im}\bar{l}_{1}$ and
$\operatorname{Im}\bar{l}_{2}$ form a proper direct sum, we have
that $\ord$ $\bar{l}y=\min\left\{ \ord\bar{l}_{1}y_{1},\ord\bar{l}_{2}y_{2}\right\} =\min\left\{ \ord y_{1},\ord y_{2}\right\} $,
where the last equality follows from the order preserving property
of $\bar{l}_{1}$ and $\bar{l}_{2}$. Using Corollary 4.8 again, together
with the fact that $\operatorname{Im}\bar{f}_{1}$ and $\mathscr{R}_{1}$
form a proper direct sum, gives that $\min\left\{ \ord y_{1},\ord y_{2}\right\} =\operatorname{ord}y$
whence $\ord$ $\bar{l}y=\operatorname{ord}y$ as claimed and the
proof is complete.~~~$\square$

Clearly, the bicausal $\Lambda K$ -linear map $\bar{l}$ of Corollary
5.7 can be regarded as a bicausal postcompensator for $\bar{f}_{1},$
and there is a kind of duality between feedback and compensation which
deserves some further comments.

Let $\bar{f}:\Lambda U\rightarrow\Lambda Y$ be an extended linear
$\mathrm{i}/\mathrm{o}$ map and let $\bar{l}_{\mathrm{pr}}:\Lambda U\rightarrow\Lambda U$
be a bicausal precompensator for $\bar{f}$. If $\bar{w}:\Lambda U\rightarrow\Lambda U$
is the strictly causal part of $\bar{l}_{\mathrm{pr}}$, then the
causal feedback problem is that of existence of a causal $\Lambda K$
-linear map $\bar{g}:\Lambda Y\rightarrow\Lambda U$ such that $\bar{w}=\bar{g}\cdot\bar{f}.$
The map $\bar{g}$ can be regarded essentially as a causal (but not
necessarily bicausal) postcompensator for $\bar{f}$. Conversely,
if $\bar{l}_{\mathrm{po}}:\Lambda Y\rightarrow\Lambda Y$ is a bicausal
postcompensator and if $\bar{w}:\Lambda Y\rightarrow\Lambda Y$ the
strictly causal part of $\bar{l}_{\mathrm{po}},$ the dual of the
above causal factorization problem is that of the existence of a causal
$\Lambda K$ -linear map $\bar{g}:\Lambda Y\rightarrow\Lambda U$
such that $\bar{w}=\bar{f}\cdot\bar{g}$. Here $\bar{g}_{-}$ can
be viewed as a causal, but again not necessarily bicausal, precompensator
for $\bar{f}$. Thus the pre- and postcompensator problems become
interrelated through feedback. We can also write down the dual of
Corollary 5.7 regarding the problem of bicausal precompensation.
\begin{cor*}
COROLLARY 5.8. Let $\bar{f}_{1},\bar{f}_{2}:\Lambda U\rightarrow\Lambda Y$
be two extended linear i/o maps with U and Y finite dimensional K-linear
spaces. There exists a bicausal $\Lambda K$ -linearmap $\bar{l}:\Lambda U\rightarrow\Lambda U$
such that $\bar{f}_{2}=\bar{f}_{1}\cdot\bar{l}$ if and only if $\ker$
$\pi^{-}\bar{f}_{1}^{*}=\operatorname{ker}\pi^{-}\bar{f}_{2}^{*},$
where $\bar{f}_{1}^{*}$ and $\bar{f}_{2}^{*}$ denote the dual maps
of $\bar{f}_{1}$ and $\bar{f}_{2}$ respectively.
\end{cor*}
In Corollary 5.8 the dual maps $\bar{f}_{1}^{*}$ and $\bar{f}_{2}^{*}$
can of course be identified with the transposes of the corresponding
maps (or transfer functions) in view of the finite dimensionality
of the underlying spaces.

In Hautus and Heymann {[}1978{]}, the static state feedback problem
was investigated. This is the following problem: Given an extended
linear i/s map $\bar{f}:\Lambda U\rightarrow\Lambda Y$ under what
conditions can a bicausal precompensator $\bar{l}:\Lambda U\rightarrow\Lambda U$
be written as $\bar{l}^{-1}=L+G\bar{f},$ where $L$ and $G$ are
static maps. It was shown there that a necessary and sufficient condition
for the static state feedback problem to have a solution is that
\[
(5.9)\bar{l}^{-1}(\operatorname{ker}\tilde{f})\subset\Omega^{+}U,
\]
where $\tilde{f}:\Omega^{+}U\rightarrow\Gamma^{+}Y$ is the restricted
i/s map associated with $\bar{f}$. We now turn to the more general
question of static output (rather than state) feedback. As we have
been doing throughout this paper, we focus our attention on the static
factorization problem which is characterized in the following
\begin{thm*}
5.10. Let $\bar{f}:\Lambda U\rightarrow\Lambda Y$ and $\bar{h}:\Lambda U\rightarrow\Lambda W$
be $\Lambda K$ -linear maps. There exists a static $\Lambda K$ -linear
map $G:\Lambda Y\rightarrow\Lambda W$ such that $\bar{h}=G\cdot\bar{f}$
if and only if $\operatorname{ker}\bar{p}_{1}\cdot\bar{f}\subset$
$\ker$ $\bar{p}_{1}\cdot\bar{h}$.
\end{thm*}
Proof. Assume first that $G$ exists so that $\bar{h}=G\cdot\bar{f}$.
Then $u\in$ $\ker$ $\bar{p}_{1}\cdot\bar{f}$ implies that $\bar{p}_{1}\cdot\bar{f}(u)=0,$
whence $\bar{p}_{1}\cdot\bar{h}(u)=\bar{p}_{1}\cdot G\cdot\bar{f}(u)=G\cdot\bar{p}_{1}\cdot\bar{f}(u)=0,$
so that $u\in\operatorname{ker}\bar{p}_{1}\cdot\bar{h}$ Conversely,
assume that $\ker$ $\bar{p}_{1}\cdot\bar{f}\subset\operatorname{ker}\bar{p}_{1}\cdot\bar{h}$.
This implies the existence of a $K$ -linear $\left.\operatorname{map}G:Y\rightarrow W\text{ such that }\bar{p}_{1}\cdot\hbar=G_{-}\cdot\bar{p}_{1}\cdot\bar{f}.\text{ By definition of static maps (see }(2.18)\right),$
we have that $G\cdot\bar{p}_{1}=\bar{p}_{1}\cdot G$ so that $\bar{p}_{1}(\bar{h}-G\cdot\bar{f})=0.$
That this implies $\bar{h}=G\bar{f}=0$ is seen as follows. Suppose
to the contrary that $(\bar{h}-G\cdot\bar{f})(u)=\sum_{t\in Z}y_{l}z^{-t}\neq0$
for some $u\in\Lambda U$ Then there exists $k\in\mathbb{Z}$ such
that $y_{k}\neq0.$ Let $\hat{u}=z^{k-1}u$ and note that $p_{1}(\bar{h}-G\bar{f})(\hat{u})=$
$p_{1}\sum_{\epsilon\in\mathbf{Z}}y_{i}z^{-i+k-1}=y_{k}\neq0,$ a
contradiction.~~~~$\square$

We shall conclude the present discussion by specializing our static
factorization results to the case of linear i/s maps. We need the
following lemma.
\begin{lem*}
$5.11.$ Let $\bar{f}:\Lambda U\rightarrow\Lambda Y$ be an injective
extended linear i/s map. Then $\operatorname{ker}\pi^{+}\bar{f}\subset\Omega^{+}U$.
\end{lem*}
Proof. Let $u\in\operatorname{ker}\pi^{+}\bar{f}$ be any element.
Then $\bar{f}(u)\in\Omega^{+}Y$ so that $\bar{p}_{1}\cdot\bar{f}(u)=0$
Write $u=u^{+}+u^{-},$ where $u^{+}\in\Omega^{+}U$ and $u^{-}\in z^{-1}\Omega^{-}U.$
Then by the strict causality of it follows that $\bar{f}\left(u^{-}\right)\in z^{-2}\Omega^{-}Y$
and $\bar{p}_{1}\cdot\bar{f}\left(u^{-}\right)=0$. Hence $\bar{p}_{1}\cdot\bar{f}\left(u^{+}\right)=$
$\bar{p}_{1}\cdot\bar{f}(u)-\bar{p}_{1}\cdot\bar{f}\left(u^{-}\right)=0$
and $u_{-}^{+}\in\operatorname{ker}\bar{p}_{1}\cdot\bar{f}\cdot j^{+}=\operatorname{ker}f=\operatorname{ker}\tilde{f},$
the last equality follow ing from the i/s property of $\bar{f}$.
We conclude that $\bar{f}\left(u^{+}\right)\in\Omega^{+}Y$ so that
also $\bar{f}\left(u^{-}\right)=$ $\bar{f}(u)-\bar{f}\left(u^{+}\right)\in\Omega^{+}Y.$
Hence $\bar{f}\left(u^{-}\right)\in\Omega^{+}Y\cap z^{-2}\Omega^{-}Y=0$
and, by the injectivity of $\bar{f},u^{-}=$ 0 concluding the proof.
$\square$ 
\begin{cor*}
5.12. Let $\bar{f}:\Lambda U\rightarrow\Lambda Y$ be an injective
extended linear i/s map and let $\bar{h}:\Lambda U\rightarrow\Lambda W$
be a strictly causal $\Lambda K$ -linear map. Then there exists a
static map $G:$ $\Lambda Y\rightarrow\Lambda W$ such that $\bar{h}=G\cdot\bar{f}$
if and only if $\ker$ $\pi^{+}\bar{f}\subset\mathrm{\ker}\pi^{+}\bar{h}$.
\end{cor*}
Proof. If $G$ exists such that $\bar{h}=G\cdot\bar{f},$ then $u\in\operatorname{ker}\pi^{+}\bar{f}$
implies that $\bar{f}(u)\in\Omega^{+}Y,$ so that $\bar{h}(u)=G\cdot\bar{f}(u)\in\Omega^{+}W$
and $u\in\operatorname{ker}\pi^{+}\hbar.$ Conversely, suppose $\ker$
$\pi^{+}\bar{f}\subset$ $\ker$ $\pi^{+}\bar{h}$ We will show that
this implies that $\ker$ $\bar{p}_{1}\cdot\bar{f}\subset$ $\ker$
$\bar{p}_{1}\cdot\bar{h},$ from which the existence of $G$ is insured
by Theorem 5.10 . Let $u\in\operatorname{ker}\bar{p}_{1}\cdot\bar{f}$
be any element and write $u=u^{+}+u^{-}$ where $u^{+}\in\Omega^{+}U$
and $u^{-}\in z^{-1}\Omega^{-}U.$ Then, by strict causality of both
$\bar{f}$ and $\bar{h}$ it follows that $\bar{f}\left(u^{-}\right)\in z^{-2}\Omega^{-}Y$
and $\bar{h}\left(u^{-}\right)\in z^{-2}\Omega^{-}W$ yielding $\bar{p}_{1}\bar{f}\left(u^{-}\right)=0$
and $\bar{p}_{1}\bar{h}\left(u^{-}\right)=0$ Hence, $u^{+}=u-u^{-}\in\operatorname{ker}\bar{p}_{1}\bar{f}$
so that $u^{+}\in\operatorname{ker}f=\operatorname{ker}\hat{f},$
the last equality following from the i/s property of $\bar{f}$. Consequently
$u^{+}\in\operatorname{ker}\tilde{f}\subset\operatorname{ker}\pi^{+}\bar{f}\subset\operatorname{ker}\pi^{+}\bar{h}\subset\operatorname{ker}\bar{p}_{1}\bar{h},$
the last inclusion holding by definition. Thus $u=u^{+}+u^{-}\in\operatorname{ker}\bar{p}_{1}\bar{h},$
and the proof is complete.~~~~$\square$

Let $\bar{f}:\Lambda U\rightarrow\Lambda Y$ be a reachable linear
i/s map. Let $\bar{l}:\Lambda U\rightarrow\Lambda U$ be a bicausal
isomorphism and write $\bar{l}^{-1}=L+h,$ where $L$ is static and
$\bar{h}$ is strictly causal. Corollary 5.12 can then be interpreted
as a solvability condition of the static state feedback problem. Clearly,
the condition of the corollary must be equivalent with condition (5.9)
which was obtained in Hautus and Heymann {[}1978{]}. We shall see
next (Theorem 5.14 below) that this is indeed the case. We require
the following lemma.
\begin{lem*}
5.13. Let $\bar{f}:\Lambda U\rightarrow\Lambda Y$ be an extended
linear i/s map and let $\bar{h}:\Lambda U\rightarrow\Lambda W$ be
a strictly causal $\Lambda K$ -linear map. Then $\ker$ $\tilde{f}\subset$
$\ker$ $\tilde{h}$ only if $\ker$ $\bar{f}\subset$ $\ker$ $\bar{h}$
\end{lem*}
Proof. Assume that $\ker$ $\bar{f}\not\subset$ $\ker$ $\bar{h}$
and let $u\in\operatorname{ker}\bar{f}$ satisfy $\bar{h}(u)\neq0.$
Then there exists $k\in\mathbb{Z}$ such that $\pi^{+}\bar{h}\left(z^{k}u\right)\neq0$
so that by the strict causality of $\bar{h}$ we have that $0\neq\mathscr{S}^{+}\left(z^{k}u\right)\in\Omega^{+}U$
and $\pi^{+}\bar{h}\left(\mathscr{S}^{+}\left(z^{k}u\right)\right)=\tilde{h}\left(\mathscr{S}^{+}\left(z^{k}u\right)\right)\neq0.$
However, $\bar{f}\left(z^{k}u\right)=0$ and upon application of Proposition
5.6 we also have that $\bar{f}\left(\mathscr{S}^{+}\left(z^{k}u\right)\right)=0,$
whence $\mathscr{S}^{+}\left(z^{k}u\right)\in$ $\ker$ $\tilde{f}$.
Thus $\ker$ $\hat{f}\notin$ $\ker$ $\tilde{h}$ and the proof is
complete. $\square$ 
\begin{thm*}
$5.14.$ Let $\bar{f}:\Lambda U\rightarrow\Lambda Y$ be a reachable
extended linear i/s map. Let $\bar{l}:\Lambda U\rightarrow\Lambda U$
be a bicausal $\Lambda K$ -linear map and write $\bar{l}^{-1}=L+\bar{h}$
where $L$ is static and $\bar{h}$ is strictly causal. Then $\operatorname{ker}\pi^{+}\bar{f}\subset\operatorname{ker}\pi^{+}\bar{h}$
if and only if $\bar{l}^{-1}(\operatorname{ker}\tilde{f})\subset\Omega^{+}U$.
\end{thm*}
Proof. Suppose $\ker$ $\pi^{+}\bar{f}\subset$ $\ker$ $\pi^{+}\bar{h}$.
Let $u\in$ $\ker$ $\tilde{f}$ be any element. Then $u\in$ $\ker$
$\pi^{+}\bar{h},$ and since $u\in\Omega^{+}U$ we also have that
$u\in\operatorname{ker}\pi^{+}L.$ Hence $u\in$ $\left(\text{ \ensuremath{\ker\ }}\pi^{+}\bar{h}\right)\cap\left(\text{ \ensuremath{\ker\ }}\pi^{+}L\right)\subset\pi^{+}(\bar{h}+L)=\operatorname{ker}\pi^{+}\bar{l}^{-1}$
so that $\bar{l}^{-1}(u)\in\Omega^{+}U.$ Conversely assume that $\bar{l}^{-1}(\operatorname{ker}\tilde{f})\subset\Omega^{+}U.$
This immediately implies that $\ker$ $\tilde{f}\subset$ $\ker$
$\tilde{h}$ whence, by Lemma $5.13,$ $\ker$ $\bar{f}\subset$ $\ker$
$\bar{h}$. Now let $u\in$ $\ker$ $\pi^{+}\bar{f}$ and write $u=u^{+}+u^{-}$
with $u^{+}\in\Omega^{+}U$ and $u^{-}\in z^{-1}\Omega^{-}U.$ Then
$\bar{f}\left(u^{-}\right)\in z^{-2}\Omega^{-}Y,$ and since $\bar{f}(u)\in\Omega^{+}Y$
we conclude that $\bar{p}_{1}\cdot\bar{f}\left(u^{+}\right)=0.$ This
implies that $u^{+}\in\operatorname{ker}f=\operatorname{ker}\tilde{f}$
(with the equality holding since $\bar{f}$ is an i/s map) so that
$u^{+}\in\operatorname{ker}\tilde{h}\subset$ $\ker$ $\pi^{+}\bar{h}$.
Finally, $u^{+}\in$ $\ker$ $\tilde{f}$ implies that $\bar{f}\left(u^{+}\right)\in\Omega^{+}U$
whence $\bar{f}\left(u^{-}\right)=\bar{f}(u)-\bar{f}\left(u^{+}\right)\in\Omega^{+}Y.$
But then $\bar{f}\left(u^{-}\right)\in\Omega^{+}Y\cap z^{-2}\Omega^{-}Y=0,$
so that $u^{-}\in\operatorname{ker}\bar{f}\subset\operatorname{ker}\bar{h},$
and hence $u^{-}\in\operatorname{ker}\pi^{+}\bar{h}.$ This implies
that $u=u^{+}+u^{-}\in\operatorname{ker}\pi^{+}\bar{h}$ concluding
the proof.~~~~$\square$

\section*{6. Factorization invariants- explicit calculation.}

Throughout this section we shall assume that $U=K^{m}$ and $Y=K^{p}$,
and we shall study properties of $\Lambda U$ as an $\Omega^{-}K$
-module as well as properties of submodules thereof

The ring $\Omega^{-}K$ is of course a principal ideal domain, and
clearly also a Euclidean domain. The units of $\Omega^{-}K$ are precisely
those elements whose order is zero and each element $0\neq\alpha\in\Omega^{-}K$
can be expressed as
\[
\alpha=z^{-\ord\alpha}\alpha_{0}
\]
where $\alpha_{0}\in\Omega^{-}K$ is a unit. It is clear, therefore,
that all the ideals of $\Omega^{-}K$ are of the form $\left(z^{-k}\right),$
forming a chain with $\left(z^{-1}\right)$ being the unique maximal
ideal and the only prime. Thus, the ring $\Omega^{-}K$ is also a
local ring and $\Omega^{-}K/\left(z^{-1}\right)$ is a field, isomorphic
to the field $\mathscr{K}_{0}$ which consists of the units of $\Omega^{-}K$
augmented by zero. We shall make use of the special properties of
the ring $\Omega^{-}K$ in the ensuing discussion.

For a fixed integer $k$, consider the subset $z^{-k}\Omega^{-}U\subset\Lambda U$.
Clearly, this subset is an $\Omega^{-}K$ submodule of $\Lambda U.$
Moreover, while $\Lambda U$ itself is not a finitely generated $\Omega^{-}K$
module, the submodule $z^{-k}\Omega^{-}U$ is (and hence is a free
module). In fact, it is readily noted that $\operatorname{rank}_{\Omega^{-}K}z^{-k}\Omega^{-}U=\operatorname{dim}_{\Lambda K}\Lambda U=\operatorname{dim}_{K}U.$
Indeed, if $\left\{ e_{1},\cdots,e_{m}\right\} $ is a basis for $U$
(as well as for $\Lambda U$ ), then $\left\{ z^{-k}e_{1},\cdots,z^{-k}e_{m}\right\} $
is a basis (i.e., a free generator) for $z^{-k}\Omega^{-}U$.

Let $0\neq\Delta\subset\Lambda U$ be an $\Omega^{-}K$ -submodule.
We say that $\Delta$ is of finite order if there exists a finite
integer $k$ such that $\Delta\subset z^{-k}\Omega^{-}U.$ The maximal
integer $k$ for which the above holds, and which is the least order
of elements in $\Delta$, is denoted $k_{\Delta}$ and is called the
order of $\Delta.$ We define the order of the zero module as infinity.
We have the following:
\begin{prop*}
6.1. Let $0\neq\Delta\subset\Lambda U$ be an $\Omega^{-}K$ -submodule.
Then $\Delta$ is finitely generated if and only if it has finite
order.
\end{prop*}
Proof. If $\Delta$ has finite order there exists a finite integer
$k$ such that $\Delta$ is a submodule of $z^{-k}\Omega^{-}U$ which
is, of course, finitely generated. since $\Omega^{-}K$ is a principal
ideal domain, $\Delta$ is then also finitely generated. Conversely,
if $\Delta$ is finitely generated, say by elements $d_{1},\cdots,d_{m}\in\Delta,$
then clearly $\Delta\subset z^{-k_{2}}\Omega^{-}U,$ where $k_{\Delta}:=\min\left\{ \ord d_{i},i=\right.$
$1,\cdots,m\}$.~~~~~$\square$

Let $\Delta\subset\Lambda U$ be a finitely generated $\Omega^{-}K$
-submodule. Then, by Proposition 6.1 , it is of finite order and hence
rank $\Delta\leqq\operatorname{dim}U(=m).$ Let $\Delta$ be of rank
$n$ and let $d_{1},\cdots,d_{n}$ be a basis for $\Delta$. Define
the $\Omega^{-}K$ -homomorphism $D:\Omega^{-}K^{n}\rightarrow\Delta$
by $De_{i}=d_{i},i=$ $1,\cdots,n,$ where $e_{1},\cdots,e_{n}$ denotes
the natural basis for $K^{n}\left(\text{ as well as for }\Omega^{-}K^{n}\right).$
We can view $D$ also as a matrix with entries in $\Lambda K$ by
regarding $d_{i}\in\Lambda K^{m}(=\Lambda U)$ as the $i$th column
of $D$. Conversely, if $D$ is an $m\times n$ matrix with entries
in $\Lambda K$, we can regard $D$ as an $\Omega^{-}K$ -homomorphism
$\Omega^{-}K^{n}\rightarrow\Lambda U:e_{i}\mapsto d_{i},i=1,\cdots,n,$
where $d_{i}\in\Lambda U$ is the $i$th column of $D$. The image
$\Delta=D\Omega^{-}K^{n}:=\left\{ Dw|w\in\Omega^{-}K^{n}\right\} $
is an $\Omega^{-}K$ -submodule of $\Lambda U$ Clearly, rank $\Delta=\operatorname{rank}D,$
where rank $D$ is the matrix rank of $D$ over the ring $\Omega^{-}K$
(or over $\Lambda K)$.

Consider now the special case when $n=m$ (that is, $K^{m}=U$ ) and
let $D$ be a nonsingular $m\times m$ matrix with entries in $\Lambda K$.
Then $D$ defines, as above, an $\Omega^{-}K$ homomorphism $\Omega^{-}U\rightarrow\Lambda U$
and also (when simply regarded as a transfer function) a $\Delta K$
-linear map $\Lambda U\rightarrow\Lambda U.$ Denoting both maps by
the same symbol $D,$ it is readily verified that the diagram in Fig.
6.1 is commutative,

\begin{figure}[h]
\begin{centering}
\includegraphics{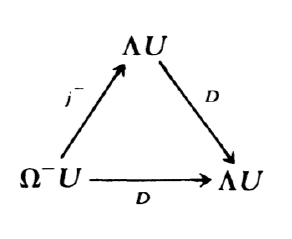}
\par\end{centering}
\caption{FIG. 6.1}

\end{figure}
where $j^{-}$ denotes the canonical injection. since the matrix $D$
is nonsingular, the $\Lambda K$ -linear map $D$ is invertible. We
shall say that the matrix $D$ is bicausal if the associated $\Lambda K$
-linear map is bicausal, i.e., if the entries of $D$ are in $\Omega^{-}K$
and its determinant is a unit in this ring (that is, has order zero).
In analogy we shall say that a matrix $D$ is strictly causal or causal
if so is the associated $\Lambda K$ -linear map. Finally, an $\Omega^{-}K$
-submodule $\Delta=D\Omega^{-}U\subset\Lambda U$ is called a full
submodule if rank $\Delta=m,$ i.e., if the matrix $D$ is nonsingular.
\begin{thm*}
6.2. Let $\Delta_{1},\Delta_{2}\subset\Lambda U$ be finitely generated
$\Omega^{-}K$ -submodules given by $\Delta_{1}=D_{1}\Omega^{-}U$
and $\Delta_{2}=D_{2}\Omega^{-}U.$ Then $\Delta_{2}\subset\Delta_{1}$
if and only if there exists a causal matrix $R$ (i.e., with entries
in $\Omega^{-}K$ ) such that $D_{2}=D_{1}R$.
\end{thm*}
The proof of Theorem 6.2 is elementary and will be omitted. The following
corollary will be useful in the sequel.
\begin{cor*}
6.3. Let $\Delta_{1},\Delta_{2}\subset\Lambda U$ be finitely generated
$\Omega^{-}K$ -submodules given by $\Delta_{1}=D_{1}\Omega^{-}U$
and $\Delta_{2}=D_{2}\Omega^{-}U.$ Assume that $\Delta_{1}$ is full
and define $R:=D_{1}^{-1}D_{2}.$ Then $\Delta_{2}\subseteq\Delta_{1}$
if and only if $R$ is causal with equality if and only if $R$ is
bicausal.
\end{cor*}
Let $\Delta\subset\Lambda U$ be a finitely generated $\Omega^{-}K$
-submodule of rank $n$ and order $k_{\Delta}$. Then for all integers
$j\leqq k_{\Delta},\Delta\subset z^{-j}\Omega^{-}U$ and for each
integer $j\geqq k_{\Delta}$ we define the submodule $\Delta_{j}\subset\Delta$
by 
\[
(6.4)~~~~~\Delta_{i}:=\Delta\cap z^{-i}\Omega^{-}U.
\]
Clearly $z^{-i}\Omega^{-}U\subset z^{-k}\Omega^{-}U$ for all $j\geqq k,$
and it follows that 
\[
(6.5)\quad\Delta=\Delta_{k_{\Delta}}\supset\Delta_{k_{2}+1}\supset\cdots>\Delta_{i}\supset\Delta_{j+1}\cdots
\]
As an immediate consequence of the fact that if $u\in\Delta_{i}$
then $z^{-1}u\in\Delta_{j+1},$ it is clear that $\operatorname{rank}\Delta=\operatorname{rank}\Delta_{j}$
for all $j$ and the quotient modules 
\[
(6.6)~~~~~\mathscr{D}_{i}:=\Delta_{i}/\Delta_{i+1}
\]
are all torsion modules with $z^{-1}$ as annihilators, that is, for
each $j$ and for each $[u]\in\mathscr{D}_{i}$ $z^{-1}[u]=0.$ Next
we shall show that the sequence of quotient modules $\left\{ \mathscr{D}_{i}\right\} $
is isomorphic to a chain $\left\{ \boldsymbol{S}_{i}\right\} $ of
(finite dimensional) $K$ -linear subspaces of $U$, that is, each
$\mathscr{D}_{i}$ is isomorphic to a subspace $S_{i}\subset U$ and
\[
(6.7)~~~~~0=S_{k_{\Delta}-1}\subset S_{k_{\Delta}}\subset S_{k_{\Delta}+1}\subset\cdots\subset S_{j}\subset\cdots\subset U.
\]
Indeed, each element in $\mathscr{D}_{i}$ is an equivalence class
$[u]$ of elements in $\Delta_{i}$. A representative $u\in[u]$ can
be expressed as $u=\sum_{k=j}^{\infty}u_{k}z^{-k}$. If $u^{\prime}=\sum_{k=j}^{\infty}u_{k}^{\prime}z^{-k}$
and $u^{\prime\prime}=\sum_{k=j}^{\infty}u_{k}^{\prime\prime}z^{-k}$
are any two elements in the same equivalence class {[}u{]} then, since
$u^{\prime}-u^{\prime\prime}\in\Delta_{j+1},$ it follows that $u_{i}^{\prime}=u_{i}^{\prime\prime}.$
Thus, with each equivalence class $[u]$ is associated a unique leading
coefficient $u_{i}$ (of $z^{-j}$ ). We can now define the map $\gamma_{i}:\mathscr{D}_{i}\rightarrow U:[u]\mapsto u_{j}$.
Naturally the $\operatorname{map}\gamma_{i}$ is $K$ -linear since
$\gamma_{j}\left([u]+\left[u^{\prime}\right]\right)=\gamma_{i}\left(\left[u+u^{\prime}\right]\right)=u_{i}+u_{i}^{\prime}$
and $\gamma_{i}(\alpha[u])=\gamma_{j}([\alpha u])=$ $\alpha u_{j}$.
It is also clear that $\gamma_{j}$ is injective, since $\ker$ $\gamma_{j}=\Delta_{j+1}=[0].$
Now, for each integer $j$ we define $S_{i}:=\operatorname{Im}\left(\gamma_{i}\right).$
Clearly $S_{i}$ is then $K$ -linearly isomorphic to $\mathscr{D}_{i}$
and $S_{i}\subset S_{j+1}$ with $S_{k_{\Delta}-j-1}=0$ for all $j\geqq0.$
Also, by the finite dimensionality of $U$, there exists an integer
$k^{\Delta^{2}}\left(\geqq k_{\Delta}\right)$ such that $S_{k^{\Delta}-1}\neq S_{k^{\Delta}}$
and $S_{k^{\Delta}+j}=S_{k^{\Delta}}$ for all $j\geqq0.$ We call
the chain $\left\{ S_{i}\right\} $ the orderchain of $\Delta$, and
the sequence of integers $\left\{ \mu_{i}\right\} ,\mu_{i}:=\operatorname{dim}S_{j}$,
we call the order list of $\Delta$ In the special case when $\Delta=$
$\ker$ $\pi^{-}\bar{f}$ where $\bar{f}$ is a linear i/o map, we
refer to the order chain and the order list of $\Delta$, respectively,
also as the \emph{latency chain} and\emph{ latency list} of $\bar{f}$.

It is interesting to observe that the integer $k^{\Delta}$ is also
the least integer satisfying the condition that $z^{-1}\Delta_{j}=\Delta_{j+1}$
for all $j\geqq k^{\Delta}$. Indeed, we have seen that $z^{-1}/j_{j}\subset\Delta_{j+1}$
for all j. To see that $z^{-1}\Delta_{j}\supset\Delta_{j+1}$ if and
only if $j\geqq k^{\Delta}$, let $u=\sum_{k=j+1}^{\infty}u_{kj}z^{-k}\in\Delta_{j+1}$
be any element. Then we can write $u=z^{-1}u^{\prime}$ where $u^{\prime}=\sum_{k=j}^{\infty}u_{k+1}z^{-k-j+1}\epsilon z^{-i}\Omega^{-}U,$
and clearly $u\in z^{-1}\Delta_{j}$ if and only if $u^{\prime}\in\Delta_{i},$
This can hold for every $u\in\Delta_{i+1}$ only if $S_{i+1}=S_{i},$
whence the necessity that $j\geqq k^{\Delta}.$ The sufficiency of
the condition is an immediate consequence of Theorem 6.11 below.

Next we have the following useful result.
\begin{lem*}
$6.8.$ Let $\Delta\subset\Lambda U$ be a finitely generated $\Omega^{-}K$
-submodule with order chain $\left\{ \boldsymbol{S}_{i}\right\} $
and order list $\left\{ \mu_{i}\right\} .$ Then $\operatorname{dim}S_{k^{\Delta}}=\operatorname{rank}\Delta$.
\end{lem*}
Proof. Let $\operatorname{rank}\Delta=\mu,$ let $d_{1},\cdots,d_{\mu}$
be a basis of $\Delta$ and define $\mathscr{R}:=\operatorname{span}_{\Lambda K}\left\{ d_{1},\cdots,d_{\mu}\right\} .$
It is easily seen that $\mathscr{R}$ is the smallest $\Lambda K$
-linear space containing $\Delta$ and $\operatorname{dim}_{\Lambda K}\mathscr{R}=\operatorname{rank}\Delta.$
The $\Lambda K$ -linear space $\mathscr{R}$ has a proper basis and
$(\mathrm{by}$ Corollary 4.5 ) $\operatorname{dim}_{\Lambda K}\mathscr{R}=\operatorname{dim}_{K}\widehat{R}.$
But clearly $\hat{\mathscr{R}}=S_{k^{2}}$ and the proof is complete.
$\square$

Let $\left\{ S_{i}\right\} $ and $\left\{ S_{i}^{\prime}\right\} $
be the order chains and $\left\{ \mu_{i}\right\} $ and $\left\{ \mu_{i}^{\prime}\right\} $
the order lists, respectively of submodules $\Delta$ and $\Delta^{\prime}$
of $\Lambda U.$ We shall say that $\left\{ S_{i}^{\prime}\right\} $
is a subchain of $\left\{ S_{i}\right\} ,$ denoted $\left\{ S_{i}^{\prime}\right\} \subset\left\{ S_{i}\right\} $
if, for all $j,S_{i}^{\prime}\subset S_{j}$. Similarly we say that
the list $\left\{ \mu_{i}^{\prime}\right\} $ is smaller than the
list $\left(\mu_{i}\right\} ,$ denoted $\left\{ \mu_{i}^{\prime}\right\} \leqq\left\{ \mu_{i}\right\} $
if $\mu_{i}^{\prime}\leqq\mu_{i}$ for all integers $j$. As an immediate
consequence of the definition we have the following,
\begin{prop*}
6.9. Let $\Delta,\Delta^{\prime}\subset\Lambda U$ be $\Omega^{-}K$
-submodules with order chains $\left\{ S_{i}\right\} $ and $\left\{ S_{i}^{\prime}\right\} $
and order lists $\left\{ \mu_{i}\right\} $ and $\left\{ \mu_{i}^{\prime}\right\} ,$
respectively. If $\Delta^{\prime}\subset\Delta$ then $\left\{ S_{i}^{\prime}\right\} \subset\left\{ S_{i}\right\} $
and $\left\{ \mu_{i}^{\prime}\right\} \leqq\left\{ \mu_{i}\right\} $. 
\end{prop*}
Let $\Delta\subset\Lambda U$ be a finitely generated $\Omega^{-}K$
-submodule. A set of elements $d_{1},\cdots,d_{k}\in\Delta$ is called
properly free if the elements are properly independent as elements
of $\Lambda U$ (regarded as a $\Lambda K$ -linear space), that is,
if the leading coefficients $\hat{d}_{1},\cdots,\hat{d}_{k}$ are
$K$ -linearly independent. It is then clear that if $d_{1},\cdots,d_{k}$
are properly free they are also free (i.e. independent over the ring
$\left.\Omega^{-}K\right)$.
\begin{defn*}
6.10. Let $\Delta\subset\Lambda U$ be a finitely generated $\Omega^{-}K$
-submodule. A basis $d_{1},\cdots,d_{\mu}$ of $\Delta$ is called
proper if $d_{1},\cdots,d_{\mu}$ are properly free. The basis will
be called ordered if $\ord$ $d_{i+1}\geqq$ $\ord$ $d_{i}$ for
all $i=1,\cdots,\mu-1$.
\end{defn*}
\begin{thm*}
$6.11.$ Let $\Delta\subset\Lambda U$ be an $\Omega^{-}K$ -submodule
or rank $\mu$ and of order $k_{\Delta},$ with order chain $\left\{ \boldsymbol{S}_{i}\right\} $
and order list $\left\{ \mu_{i}\right\} .$ Then (i) there exists
an ordered proper basis for $\Delta$. (ii) If $d_{1},\cdots,d_{\mu}$
is any ordered proper basis for $\Delta$, then the following conditions
are satisfied:
\begin{lyxlist}{MMM}
\item [{\textup{(6.12)}}] $\operatorname{ord}d_{i}=i\quad for\mu_{i-1}<j\leqq\mu_{i}andi=k_{\Delta},k_{\Delta+1},\cdots$ 
\item [{(6.13)}] For each $j=1,\cdots,\mu,$ the set $\hat{d}_{1},\cdots,\hat{d}_{j}\in S_{i},$
where i is the least integer such that $j\leqq\mu_{i}$.
\end{lyxlist}
\end{thm*}
Proof. (i) We shall construct an ordered proper basis for $\Delta$
which, in particular, satisfies (6.12) and $(6.13).$ Consider the
sequence $\left\{ \mathscr{D}_{i}\right\} $ of quotient modules $\mathscr{D}_{j}$
defined by $(6.6),$ of which $\mathscr{D}_{k_{\Delta}}$ is the first
nonzero one. Choose any equivalence class $0\neq\left[d_{1}\right]\epsilon$
$\mathscr{D}_{k_{\Delta}}$ and let $d_{1}\in\Delta$ be any representative
of $\left[d_{1}\right].$ Then $\ord$ $d_{1}=k_{\Delta}$ and $d_{1}$
is clearly properly free. We proceed stepwise and assume that for
$j>0,d_{1},\cdots,d_{j}$ are properly free elements of $\Delta$
satisfying (6.12) and $(6.13).$ If $j<\mu,$ let $k$ denote the
least integer such that $j<\mu_{k}.$ Then $\hat{d}_{1},\cdots,\hat{d}_{i}\in S_{k}$
are $K$ -linearly independent, but they do not span $S_{k},$ since
$\operatorname{dim}S_{k}=\mu_{k}.$ Thus, there exists an element
$\left[d_{i+1}\right]\in\mathscr{D}_{k}$ such that for any representative
$d_{j+1}\in\left[d_{j+1}\right]$, the set $\hat{d}_{1},\cdots,\hat{d}_{j},\hat{d}_{j+1}\in S_{k}$
are $K$ -linearly independent and hence the set $d_{1},\cdots,d_{i+1}$
is properly free. Clearly (6.13) is satisfied, and since $\ord$ $d_{j+1}=k$
so is also $(6.12).$ By Lemma $6.8,\operatorname{dim}S_{k^{\Delta}}=\operatorname{rank}\Delta=\mu,$
so that we finally obtain an ordered, properly free set of elements
$d_{1},\cdots,d_{\mu}\in\Delta$ satisfying (6.12) and $(6.13).$
Let $\Delta^{\prime}$ denote the $\Omega^{-}K$ -submodule of $\Lambda U$
generated by $d_{1},\cdots,d_{\mu}.$ It remains to be shown that
$\Delta^{\prime}=\Delta.$ Obviously $\Delta^{\prime}\subset\Delta$
and since $\ord$ $d_{i}\leqq k^{\Delta}$ for all $i=1,\cdots,\mu$
and $\operatorname{since}\operatorname{span}_{K}\left\{ \hat{d}_{1},\cdots,\hat{d}_{\mu}\right\} =S_{k^{\Delta}},$
it follows also that $\Delta_{k^{\Delta}}\subset\Delta^{\prime}.$
Let $u\in\Delta$ be any element and let $\ord$ $u=j.$ Then $\hat{u}\in S_{i}$
whence there are elements $\alpha_{1},\cdots,\alpha_{\mu_{i}}\in\Omega^{-}K$
such that $\sum_{k=1}^{\mu_{i}}\hat{\alpha}_{k}\hat{d}_{k}=\hat{u}$
and $\ord$ $\left(u-\sum_{k=1}^{\mu_{i}}\alpha_{k}d_{k}\right)>j.$
Proceeding stepwise the same way, we conclude that there are elements
$\alpha_{1},\cdots,\alpha_{\mu}\in\Omega^{-}K$ such that $u=\sum_{i=1}^{\mu}\alpha_{i}d_{i}+u^{\prime},$
with $\ord$ $u^{\prime}\geqq k^{\Delta}.$ Clearly, $\sum_{i=1}^{\mu}\alpha_{i}d_{i}\in\Delta^{\prime},$
and since $u^{\prime}\in\Delta_{k^{2}}\subset\Delta^{\prime},$ it
follows also that $u\in\Delta^{\prime}$ and the proof of (i) is complete.
To see that (ii) holds, it suffices to observe that for each integer
$j$, every ordered proper basis $d_{1},\cdots,d_{\mu}$ of $\Delta$
has precisely $\mu_{i}$ elements whose order is less than or equal
to $j$ and $\operatorname{span}_{K}\left\{ \hat{d}_{1},\cdots,\hat{d}_{\mu}\right\} =S_{j}$.~~~~~$\square$

The following immediate corollary to Theorem 6.11 gives a sharp insight
to the relation between ordered proper bases of $\Omega^{-}K$ -modules
and their order chain
\begin{cor*}
6.14. Let $\Delta\subset\Lambda U$ be an $\Omega^{-}K$ -submodule
of rank $\mu$ with order chain $\left\{ S_{i}\right\} $ and order
list $\left\{ \mu_{i}\right\} .$ Then $d_{1},\cdots,d_{\mu}$ is
an ordered proper basis of $\Delta$ if and only if for each $j,\hat{d}_{1},\cdots,\hat{d}_{\mu_{i}}$
is a basis for $S_{i}$.
\end{cor*}
We now return to questions connected with our primary objective of
studying causal factorization and feedback. First we have some preliminary
facts
\begin{lem*}
$6.15.$ Let $U$ be an $m$ -dimensional K-linear space and let $\bar{f}:\Lambda U\rightarrow\Lambda Y$
be a MK-linear map. For each integer $j$ let $\Delta_{j}(\bar{f})$
the $\Omega^{-}K$-subodule of $\Lambda U$ defined by $\Delta_{j}(\bar{f}):=\operatorname{ker}\pi^{-}\bar{f}\cap z^{-j}\Omega^{-}U$.
Then rank $\Delta_{j}(\bar{f})=m$.
\end{lem*}
Proof. First note that since $\Delta_{j}(\bar{f})\subset z^{-j}\Omega^{-}U,$
rank $\Delta_{j}(\bar{f})\leqq m,$ with equality obviously holding
when $\bar{f}=0,$ since then $\ker$ $\pi^{-}\bar{f}=\Lambda U$.
Assume now that $\bar{f}\neq0$, define $t:=\max\{j-\ord\bar{f},-\ord\bar{f}\}$
and let $u\in z^{-1}\Omega^{-}U_{-}$ be any element. Then $\ord$
$\bar{f}u\geqq$ $\operatorname{ord}\bar{f}+\operatorname{ord}u\geqq\operatorname{ord}\bar{f}+t\geqq\max\{j,0\}$
and $u\in\Delta_{j}(\bar{f}).$ Hence $z^{-1}\Omega^{-}U\subset\Delta_{i}(\bar{f})$
so that $\operatorname{rank}\Delta_{j}(\bar{f})\geqq m$ and the proof
is complete. $\square$
\begin{prop*}
6.16. Let U be an m-dimensional K-linear space and let $\bar{f}:\Lambda U\rightarrow\Lambda Y$
be a $\Lambda K$ -linear map. Then the following are equivalent
\begin{lyxlist}{MM}
\item [{(i)}] $\bar{f}$ is injective.
\item [{(ii)}] $\ker$ $\pi^{-1}\bar{f}$ is finitely generated.
\item [{(iii)}] rank $\ker$ $\pi^{-}\bar{f}=m$.
\end{lyxlist}
\end{prop*}
Proof. That (ii) and (iii) are equivalent follows immediately from
Lemma 6.15 and the fact that if $\ker$ $\pi^{-1}\bar{f}$ is finitely
generated it is of finite order, say $t,$ so that $\ker$ $\pi^{-}\bar{f}=\Delta_{t}(\bar{f}).$
To see that (ii) implies (i), recall that $\ker$ $\bar{f}\subset$
$\ker$ $\pi^{-}\bar{f}$ so that if $\ker$ $\bar{f}\neq0$ then
$\ker$ $\pi^{-}\bar{f}$ is not of finite order and hence is not
finitely generated. It remains to be shown that (i) implies (ii).
Assume that (i) holds, let $y_{1},\cdots,y_{m}$ be a normalized proper
basis for $\operatorname{Im}\bar{f}\subset\Lambda Y$ and let $u_{1},\cdots,u_{m}$
be the (unique) elements of $\Lambda U$ satisfying $\bar{f}\left(u_{i}\right)=y_{i},i=1,\cdots,m.$
The proof will be complete upon showing that $\ker$ $\pi^{-}\bar{f}$
is of finite order and, in fact, we claim that $\ker$ $\pi^{-}\bar{f}\subset z^{-t}\Omega^{-}U$
where $t:=\min\left\{ \ord u_{i}|i=1,\cdots,m\right\} $. Indeed,
if $u\in\operatorname{ker}\pi^{-}\bar{f},$ then $\bar{f}(u)\in\Omega^{-}Y$
and there are elements $\alpha_{1},\cdots,\alpha_{m}\in\Omega^{-}K$
such that $\bar{f}(u)=\sum_{i=1}^{m^{\prime}}\alpha_{i}y_{i}=\sum_{i=1}^{m}\alpha_{i}\bar{f}\left(u_{i}\right)=\bar{f}\left(\sum_{i=1}^{m}\alpha_{i}u_{i}\right)$
whence $u=\sum_{i=1}^{m}\alpha_{i}u_{i}$ so that $\ord$ $u\geqq t$.~~~~~$\square$

In view of Proposition $6.16,$ it follows that the latency kernel
of a given linear i/o map $\bar{f}$ is finitely generated if and
only if $\bar{f}$ is injective, the case which receives, of course,
most of our attention. Before proceeding further, a remark on the
noninjective case is in order
\begin{rem*}
$6.17.$ It is readily noted that if $\bar{f}:\Lambda U\rightarrow\Lambda Y$
is a $\Lambda K$ -linear map, then $\ker$ $\pi^{-}\bar{f}$ can
(always) be written as 
\[
\ker\pi^{-}\bar{f}=\operatorname{ker}\bar{f}+\mathscr{R}
\]
where $\mathscr{R}$ is a finitely generated full $\Omega^{-}K$ -submodule
of $\Lambda U$. However, in the above representation, $\mathscr{R}$
is nonunique except in the special case when $\bar{f}$ is injective
and $\ker$ $\bar{f}=0.$ If $\bar{f}_{1}$ and $\bar{f}_{2}$ are
two $\Lambda K$ -linear maps then $\ker$ $\pi^{-}\bar{f}_{1}\subset$
$\ker$ $\pi^{-}\bar{f}_{2}$ if and only if $\ker$ $\bar{f}_{1}+\mathscr{R}_{1}\subset\operatorname{ker}\bar{f}_{2}+\mathscr{R}_{2}.$
While this condition necessarily implies $\ker$ $\bar{f}_{1}\subset\operatorname{ker}\bar{f}_{2},$
it cannot be claimed, except in the injective case, that $\mathscr{R}_{1}\subset\mathscr{R}_{2}$.
Hence, for computational purposes it is convenient in the noninjective
case to resort to the fact that $\ker$ $\pi^{-}\bar{f}_{1}\subset$
$\ker$ $\pi^{-}\bar{f}_{2}$ if and only if $\Delta_{i}\left(\bar{f}_{1}\right)\subset\Delta_{j}\left(\bar{f}_{2}\right)$
for all $j,$ where $\Delta_{j}\left(\bar{f}_{i}\right)$ is as defined
in Lemma 6.15 However, $\Delta_{j}\left(\bar{f}_{1}\right)\subset\Delta_{j}\left(\bar{f}_{2}\right)$
for all $j$ if and only if $\Delta_{i}\left(\bar{f}_{1}\right)\subset\Delta_{i}\left(\bar{f}_{2}\right)$
for any $j\leqq$ $\min\left\{ \ord\mathscr{R}_{1},\ord\mathscr{R}_{2}\right\} $
where $\mathscr{R}_{i},i=1,2,$ are any submodules in the corresponding
representations of $\ker$ $\pi^{-}\bar{f}_{i}$. By Lemma 6.15 both
$\Delta_{j}\left(\bar{f}_{1}\right)$ and $\Delta_{j}\left(\bar{f}_{2}\right)$
are full finitely generated $\Omega^{-}K$ -submodules of $\Lambda U$
so that the situation is thus similar to that in the injective case.~~~~~$\square$
\end{rem*}
Let $\bar{f}:\Lambda U\rightarrow\Lambda Y$ be an injective extended
linear $\mathrm{i}/\mathrm{o}$ map and let $\Delta=\mathrm{\ker}\pi^{-}\bar{f}$.
Then $\Delta=D\Omega^{-}U$ is a full, finitely generated $\Omega^{-}K$
-submodule of $\Lambda U$ and the columns $d_{1},\cdots,d_{m}$ of
the generating matrix $D$ form a basis of $\Delta.$ We shall next
establish certain properties of possible selections of the matrix
$D$.
\begin{prop*}
$6.18.$ Let $\bar{f}:\Lambda U\rightarrow\Lambda Y$ be an injective
extended linear i/o map. Write $\ker$ $\pi^{-}\bar{f}=D\Omega^{-}U.$
Then $D^{-1}$ exists and is strictly causal; i.e., the elements of
$D^{-1}$ are in $z^{-1}\Omega^{-}K$.
\end{prop*}
Proof. The existence of $D^{-1}$ follows immediately from Proposition
$6.16.$ From the strict causality of $\bar{f}$ it follows that $z\Omega^{-}U\subset$
$\ker$ $\pi^{-}\bar{f},$ whence by Theorem 6.2 there exists a causal
matrix $R$ such that $zI=DR$. Thus $D^{-1}=z^{-1}R$ and $z^{-1}R$
is clearly strictly causal. $\square$

Let $\Delta\subset\Lambda U$ be a full finitely generated $\Omega^{-}K$
-submodule and write $\Delta=D\Omega^{-}U.$ We call the columns $d_{1},\cdots,d_{m}$
of $D$ a polynomial be : is of $\Delta$ if the matrix $D$ is a
polynomial matrix, i.e., with elements in $\Omega^{+}K.$ We call
the basis a strictly polynomial basis if its elements are strict polynomials,
i.e., with elements in $z\Omega^{+}K$. If in addition $D$ is a proper
basis we call it a \emph{proper polynomial basis}, respectively, proper
strictly polynomial basis for $\Delta$.
\begin{thm*}
$6.19.$ Let $\bar{f}:\Lambda U\rightarrow\Lambda Y$ be an injective
extended linear i/o map. Then $\ker$ $\pi^{-}\bar{f}$ has a proper
strictly polynomial basis.
\end{thm*}
Proof. Let $\tilde{d}_{1},\cdots,\tilde{d}_{m}$ be a proper basis
for $\ker$ $\pi^{-}\bar{f}$ and for each $i$ write $\tilde{d}_{i}=$
$\Sigma d_{ij}\cdot z^{-i}=d_{i}+d_{i}^{-},$ where $d_{i}=\sum_{i<0}d_{ij}z^{-j}\in z\Omega^{+}U$
and $d_{i}^{-}=\Sigma_{i\geq0}d_{ij}z^{-j}\in\Omega^{-}U.$ Then $zd_{i}^{-}\in z\Omega^{-}U\subset\operatorname{ker}\pi^{-}\bar{f},$
the inclusion following from the strict causality of $\bar{f}$. Thus
there are elements $\alpha_{ij}\in\Omega^{-}K,j=1,\cdots,m,$ so that
$zd_{i}^{-}=\sum_{i=1}^{m}\alpha_{ij}\tilde{d}_{i}$ Defining the
matrices $D:=\left[d_{1},\cdots,d_{m}\right],\tilde{D}:=\left[\tilde{d}_{1},\cdots,\tilde{d}_{m}\right]$
and $A:=\left[\alpha_{ij}\right]$ we can thus write $\tilde{D}=$
$D+z^{-1}\tilde{D}A,$ or alternatively, $D=\tilde{D}\left(I-z^{-1}A\right).$
since $A$ is causal by definition of the $\alpha_{ij}$ it follows
that $\left(I-z^{-1}A\right)$ is a bicausal matrix. Consequently,
by Corollary $6.3,$ we have $\ker$ $\pi^{-}\bar{f}=\tilde{D}\Omega^{-}U=D\Omega^{-}U$
so that the columns $d_{1},\cdots,d_{m}$ of $D$ also form a proper
basis for $\ker$ $\pi^{-}\bar{f}$. That this basis is strictly polynomial
follows directly from the definition of the $d_{i}.$~~~~~$\square$

For an injective extended linear i/o map $\bar{f}$ it is convenient
to define a set of nonnegative integers, called \emph{latency indices},
which are associated in one-one correspondence with the latency list
of $\bar{f}$. We proceed as follows. Let $d_{1},\cdots,d_{m}$ be
an ordered proper basis for $\ker$ $\pi^{-}\bar{f}$. Then, as we
have seen, for each $i=1,\cdots,m$ $\ord$ $d_{i}\leqq-1.$ We define
the latency indices $\left\{ \nu_{1},\cdots,\nu_{m}\right\} $ of
$\bar{f}$ by $\nu_{i}:=-$ $\ord$ $d_{i}-1.$ The relation of the
latency indicates with the latency list is clearly established by
Corollary $6.14,$ and if $\left\{ \mu_{i}\right\} $ is the latency
list of $\bar{f}$ then we have 
\[
(6.20)~~~~~v_{i}=-j-1\text{ for }\mu_{j-1}<i\leqq\mu_{i},\quad j=k_{\Delta},k_{\Delta}+1,\cdots
\]
 where $k_{\Delta}=$ $\ord$ $\ker$ $\pi^{-}\bar{f}.$ Clearly $\nu_{i}\geqq0$
for all $i=1,\cdots,m,$ and $\bar{f}$ is nonlatent if and only if
all its latency indices are zero.

We conclude this section with the discussion of certain invariance
properties of the latency indices. We have seen previously that if
$\bar{f}_{1}:\Lambda U\rightarrow\Lambda Y$ and $\bar{f}_{2}:\Lambda U\rightarrow\Lambda Y$
are two extended linear i/o maps and if $\bar{l}_{\mathrm{po}}:\Lambda Y\rightarrow\Lambda Y$
is a $\Lambda K$ -linear bicausal isomorphism such that $\bar{f}_{2}=\bar{l}_{\mathrm{po}}\cdot\bar{f}_{1},$
then $\bar{f}_{1}$ and $\bar{f}_{2}$ have the same latency kernels;
i.e., $\ker$ $\pi^{-}\bar{f}_{1}=$ $\ker$ $\pi^{-}\bar{f}_{2}$.
If there exist both a bicausal postcompensator as above and a $\Lambda K$
-linear bicausal precompensator $\bar{l}_{\mathrm{pr}}:\Lambda U\rightarrow\Lambda U$
such that $\bar{f}_{2}=\bar{l}_{\mathrm{po}}\cdot\bar{f}_{1}\cdot\bar{l}_{\mathrm{pr}}$,
then $\ker$ $\pi^{-}\bar{f}_{2}=$ $\ker$ $\pi^{-}\bar{f}_{1}\cdot\overline{1}_{\mathrm{pr}},$
and since $u\in\operatorname{ker}\pi^{-}\bar{f}_{1}\cdot\bar{l}_{\mathrm{pr}}$
if and only if $\bar{l}_{\mathrm{pr}}u\in\operatorname{ker}\pi^{-}\bar{f}_{1},$
it follows that $\bar{l}_{\mathrm{pr}}$ $\ker$ $\pi^{-1}\bar{f}_{2}=$
$\ker$ $\pi^{-}\bar{f}_{1}.$ since the map $\bar{l}_{\mathrm{pr}}$
is, in particular, also an $\Omega^{-}K$ -homorphism (which we denote
$l_{\mathrm{pr}}$ ) we interpret it as an order preserving $\Omega^{-}K$
-isomorphism $l_{\mathrm{pr}}:\operatorname{ker}\pi^{-}\bar{f}_{2}\rightarrow\operatorname{ker}\pi^{-}\bar{f}_{1}.$
Suppose, conversely, that there exists an order preserving $\Omega^{-}K$-isomorphism
$l_{\mathrm{pr}}$ as above. Fix an integer $j$ and define (as in
Lemma 6.15) $\Delta_{j}\left(\bar{f}_{2}\right)\subset\operatorname{ker}\pi^{-}\bar{f}_{2}.$
Then, by the same lemma, $\Delta_{j}\left(\bar{f}_{2}\right)$ is
a full finitely generated $\Omega^{-}K$ submodule of $\Lambda U$,
and if $d_{1},\cdots,d_{m}$ is a proper basis for $\Delta_{i}\left(\bar{f}_{2}\right),$
it is clearly also a basis for $\Lambda U.$ Let $\bar{l}_{\mathrm{pr}}:\Lambda U\rightarrow\Lambda U$
be the (unique) $\Lambda K$ -linear map whose action on the $d_{i}^{\prime}$
s is that of $l_{\mathrm{pr}}.$ Then, $\bar{l}_{\mathrm{pr}}$ is
order preserving and thus a bicausal isomorphism $\Lambda U\rightarrow\Lambda U$
Moreover, since $\bar{l}_{\mathrm{pr}}u=l_{\mathrm{p}_{t}u}$ for
all elements $u\in\mathrm{\ker}\pi^{-}\bar{f}_{2},$ it follows that
$\bar{l}_{\mathrm{pr}}$ $\ker$ $\pi^{-}\bar{f}_{2}=$ $\ker$ $\pi^{-}\bar{f}_{1}$
whence $\ker$ $\pi^{-}\bar{f}_{2}=$ $\ker$ $\pi^{-}\bar{f}_{1}\cdot\bar{l}_{\mathrm{pr}}.$
Applying now Corollary 5.7 to the above kernel equality, we conclude
that there exists a bicausal $\Lambda K$ -linear postcompensator
$\bar{l}_{\mathrm{po}}:\Lambda Y\rightarrow\Lambda Y$ such that $\bar{f}_{2}=\bar{l}_{\mathrm{po}}\bar{f}_{1}\bar{l}_{\mathrm{pr}}.$
We have just proved the following.
\begin{thm*}
$6.21.$ Let $\bar{f}_{1},\bar{f}_{2}:\Lambda U\rightarrow\Lambda Y$
be two extended linear i/o maps with $U$ and $Y$ finite dimensional
K-linear spaces. There exist bicausal $\Lambda K$ -linear compensators
$\bar{l}_{\mathrm{pr}}:\Lambda U\rightarrow\Lambda U$ and $\bar{l}_{\mathrm{po}}:\Lambda Y\rightarrow\Lambda Y$
such that $\bar{f}_{2}=\bar{l}_{\mathrm{po}}\cdot\bar{f}_{1}\cdot\bar{l}_{\mathrm{pr}}$
if and only if there exists an order preserving $\Omega^{-}K$ -isomorphism
$l_{\mathrm{pr}}$ : $\ker$ $\pi^{-}\bar{f}_{2}\rightarrow$ $\ker$
$\pi^{-}\bar{f}_{1}$.
\end{thm*}
We now restrict Theorem 6.21 to the injective case to obtain the following
invariance characterization of the latency indices
\begin{cor*}
$6.22.$ Let $\bar{f}_{1},\bar{f}_{2}:\Lambda U\rightarrow\Lambda Y$
be two injective extended linear i/o maps with $U$ and $Y$ finite
dimensional $K$ -linear spaces. There exist bicausal $\Lambda K$
-linear compensators $\bar{l}_{\mathrm{pr}:\Lambda U}\rightarrow\Lambda U$
and $\bar{l}_{\mathrm{po}}:\Lambda Y\rightarrow\Lambda Y$ such that
$\bar{f}_{2}=\bar{l}_{\mathrm{po}}\cdot\bar{f}_{1}\cdot\bar{l}_{\mathrm{pr}}$
if and only if $\bar{f}_{1}$ and $\bar{f}_{2}$ have the same latency
indices.
\end{cor*}
Proof. By the injectivity of $\bar{f}_{1}$ and $\bar{f}_{2},$ both
$\Delta_{1}=\operatorname{ker}\pi^{-}\bar{f}_{1}$ and $\Delta_{2}=\operatorname{ker}\pi^{-}\bar{f}_{2}$
are of rank $m,$ where $m=\operatorname{dim}U,$ and in view of Theorem
6.21 it needs only to be shown that $\Delta_{1}$ and $\Delta_{2}$
have the same latency indices (or latency lists) if and only if there
exists an order preserving $\Omega^{-}K$ -isomorphism $l_{\mathrm{pr}}:\Delta_{2}\rightarrow\Delta_{1}$.
Let $d_{11},\cdots,d_{1m}$ and $d_{21},\cdots,d_{2m}$ be ordered
proper bases for $\Delta_{1}$ and $\Delta_{2}$, respectively, and
let $D_{1}$ and $D_{2}$ be the corresponding matrices. Then an order
preserving isomorphism $l_{\mathrm{pr}}:\Delta_{2}\rightarrow\Delta_{1}$
exists if and only if the matrix $D_{1}D_{2}^{-1}$ is bicausal which
is easily seen to be the case if and only if $\ord$ $d_{1j}=$ $\ord$
$d_{2j}$ for all $j=1,\cdots,m$. Employing Corollary 6.14 completes
the proof. $\square$

Theorem 6.21 and Corollary 6.22 could, of course, have been stated
for any $\Lambda K$ -linear maps and not only strictly causal ones.
The proofs did in no way depend on the causality properties of the
maps involved. Also, Corollary 6.22 could have been obtained as an
application of the existence of, so called, Smith canonical forms
for matrices over Euclidean rings (see, e.g., MacDuffee {[}1934{]}).

\section*{7. Precompensation and feedback.}

Let $\bar{f}:\Lambda U\rightarrow\Lambda Y$ be an extended linear
i/o map and let $\bar{l}:\Lambda U\rightarrow\Lambda U$ be a $\Lambda K$
-linear bicausal precompensator. Write $\bar{l}^{-1}=L+h$ where $L:\Lambda U\rightarrow\Lambda U$
is static and $\bar{h}:\Lambda U\rightarrow\Lambda U$ is strictly
causal. We have seen in $\$5$ that $\bar{l}$ can be realized by
a static precompensator (i.e., coordinate change in the input value
space and output feedback around $\tilde{f}_{-}$ (i.e., $\bar{h}=\bar{g}\cdot\bar{f}$
for causal $\Lambda K$ -linear map $\bar{g}:\Lambda Y\rightarrow$
$\Lambda U$ ) if and only if $\ker$ $\pi^{-}\bar{f}\subset$ $\ker$
$\pi^{-}\bar{h}$ (see Theorem 5.2). When $\bar{f}$ is a nonlatent
map, feedback realization as above is thus possible for every bicausal
map $\bar{l}$. In general, however, feedback realization is not possible
for every precompensator $\bar{l}$. We shall say that $\bar{l}$
has $a(\bar{v},\bar{g})$ representation if it can be expressed as
$\bar{l}=\bar{l}_{(\bar{v},\bar{\varepsilon})}=(I+\bar{g}\bar{f})^{-1}\bar{v}$
where $\bar{v}:\Lambda U\rightarrow\Lambda U$ is a bicausal isomorphism
and $\bar{g}:\Lambda Y\rightarrow\Lambda U$ is a causal $\Lambda K$
-linear map. We call the map $\bar{v}$ in the above representation
the precompensator remainder of the representation. The precompensator
$\bar{l}$ can thus be realized as feedback whenever $\bar{l}$ has
a $(\bar{v},\bar{g})$ representation with $\bar{v}=V,$ a static
map.

In general, the precompensator remainder $\bar{v}$ is dynamic and
can be represented as $\bar{v}=V+\bar{v}_{c}$ where $V$ is the static
part of $\bar{v}$ and $\bar{v}_{c}:\Lambda U\rightarrow\Lambda U$
is strictly causal, i.e., an extended linear i/o map. We recall (see,
in particular, Hautus and Heymann {[}1978{]}) that the dynamic characteristics
of $\bar{v}_{c}$ are determined by $\ker$ $\pi^{+}\bar{v}_{c}\cdot j^{+}$
which is an $\Omega^{+}K$ -submodule of $\Omega^{+}U$ and can be
represented by
\[
(7.1)~~~~~\ker\pi^{+}\bar{v}_{c}\cdot j^{+}=\operatorname{ker}\pi^{+}\bar{v}\cdot j^{+}=D\Omega^{+}U,
\]
where $D$ is a polynomial matrix whose columns form a basis for $\ker$
$\pi^{+}\bar{v}\cdot j^{+}.$ The degree $n$ of the determinant of
$D$ (when $D$ is nonsingular) is the dimension of the minimal state
space realizing $\bar{v}_{c}$. More specifically, if $D$ in (7.1)
is selected to be proper, i.e., the columns of $D$ are properly free
(in the sense that the leading coefficient vectors are $K$ -linearly
independent just as in 84 above), then the column degrees $\sigma_{i},i=$
$1,\cdots,m$ are the reachability indices of $\bar{v}_{c}$ and their
sum is $\sum_{i=1}^{m}\sigma_{i}=n=\operatorname{deg}\cdot\operatorname{det}D$.

It is of interest in selecting a $(\bar{v},\bar{g})$ pair representing
a given precompensator $\bar{l}$ to choose the representation in
such a way that the precompensator remainder $\bar{v}$ has least
dynamic order, i.e., is realizable by a state space of least possible
dimension. In this way the precompensator is realized "as much as
possible" by feedback. The following theorem provides a bound on
the dynamic order of the precompensator remainder $\bar{v}$ which
need not be exceeded in the realization of any bicausal precompensator
$\bar{l}$, and which is dependent only on the dynamic properties
(latency) of the i/o map $\bar{f}$ under consideration.
\begin{thm*}
7.2. Let $\bar{f}:\Lambda U\rightarrow\Lambda Y$ be an injective
extended linear i/o map with latency indices $\nu_{1}\geqq\cdots\geqq\nu_{m.}$
Let $\bar{l}:\Lambda U\rightarrow\Lambda U$ be a bicausal $\Lambda K$
-linear map. There exists $a(\bar{v},\bar{g})$ representation for
$\bar{I}$ such that the precompensator remainder $\bar{v}$ has (ordered)
reachability indices $\sigma_{1}\geqq\cdots\geqq\sigma_{m}$ satisfying
$\sigma_{i}\leqq\nu_{i},i=1,\cdots,m$.
\end{thm*}
\begin{rem*}
7.3. It is interesting to observe that Theorem 7.2 explicitly implies
what we have seen previously, namely, that if $\bar{f}$ is a nonlatent
i/o map, then every bicausal $\bar{l}$ can be realized as output
feedback. Indeed, if $\bar{f}$ is nonlatent, its latency indices
$\nu_{i}$ are all zero, whence by Theorem 7.2 there exists a pair
$(\bar{v},\bar{g})$ with $\bar{v}$ having reachability indices all
zero, that is, with $\bar{v}$ static.$\square$
\end{rem*}
To prove Theorem 7.2 we shall need the following lemmas.
\begin{lem*}
$7.4.$ Let $U$ be a finite dimensional $K$ -linear space and let
$\bar{v}:\Lambda U\rightarrow\Lambda U$ be $a$ bicausal $\Lambda K$
-linear isomorphism. Then $\ker$ $\pi^{+}\bar{v}\cdot j^{+}$ and
$\operatorname{ker}\pi^{+}\bar{v}^{-1}\cdot j^{+}$ have the same
lists of reachability indices.
\end{lem*}
Proof. By Hautus and Heymann {[}1978, Theorem 6.11{]} the lemma will
be proved upon showing that there exists an order-preserving $\Omega^{+}K$
-isomorphism $\ker$ $\pi^{+}\bar{v}\cdot j^{+}\rightarrow$ $\ker$
$\pi^{+}\bar{v}^{-1}\cdot j^{+}.$ We shall see that the map $\bar{v}$
itself, which is in particular also an order preserving $\Omega^{+}K$
-isomorphism, satisfies the required properties. Indeed, let $\xi\in$
ket $\pi^{+}\bar{v}\cdot j^{+}$ be any element. Then $\bar{v}\cdot j^{+}\xi=\bar{v}\xi\in\Omega^{+}U$
and since also $\xi\in\Omega^{+}U$ we have $\xi=\bar{v}^{-1}(\bar{v}\xi)=\bar{v}^{-1}j^{+}(\bar{v}\xi)\in\Omega^{+}U,$
whence $\bar{v}\xi\in\operatorname{ker}\pi^{+}\bar{v}^{-1}\cdot j^{+},$
completing the proof. $\square$

Let $\bar{f}:\Lambda U\rightarrow\Lambda Y$ be an injective extended
linear i/o map and let $d_{1},\cdots,d_{m}$ be a proper strictly
polynomial basis for $\ker$ $\pi^{-}\bar{f}$ (see Theorem 6.19 ),
and write $\ker$ $\pi^{-}\bar{f}=$ $D\Omega^{-}U$ where $D=\left[d_{1},\cdots,d_{m}\right].$
Then $z^{-1}D$ is also polynomial and the column degrees of $z^{-1}D$
are (by definition) the latency indices of $\bar{f}$. Below we shall
not distinguish sharply between maps and their transfer functions.
Let $\mathscr{S}^{-}:\Lambda U\rightarrow\Omega^{-}U:\Sigma u_{i}z^{-t}\mapsto$
$\Sigma_{t\geq0}u_{t}z^{-t}$ denote the causal truncation. Let $N:\Lambda U\rightarrow\Omega^{-}U$
be defined as the (unique) $\Lambda K$ -linear map whose transfer
function is given by 
\[
(7.5)~~~~~N:=\mathscr{S}^{-}\left(\bar{l}^{-1}D\right),
\]
and define the $\Lambda K$ -linear maps 
\[
(7.6)~~~~~\bar{\phi}:\Lambda U\rightarrow\Lambda U:u\mapsto ND^{-1}u,
\]
\[
(7.7)~~~~~\bar{v}^{-1}:=\bar{l}^{-1}-\bar{\phi}.
\]

\begin{lem*}
7.8. With $\bar{\phi}$ and $\bar{v}^{-1}$ as defined in (7.6) and
(7.7) the following hold true:
\begin{lyxlist}{MM}
\item [{(i)}] $\ker$ $\pi^{-}\bar{f}\subset\operatorname{ker}\pi^{-}\bar{\phi}$. 
\item [{(ii)}] $z^{-1}D\Omega^{+}U\subset\operatorname{ker}\pi^{+}\cdot\bar{v}^{-1}\cdot j^{+}$.
\end{lyxlist}
\end{lem*}
Proof. z(i) Let $u\in\operatorname{ker}\pi^{-}\bar{f}.$ Then $u=Dw$
for some $w\in\Omega^{-}U$ and we have $\bar{\phi}u=$ $ND^{-1}u=ND^{-1}Dw=Nw\in\Omega^{-}U$
since $N$ is a causal map, and hence $\pi^{-}\bar{\phi}u=0$ so that
$u\in\operatorname{ker}\pi^{-}\bar{\phi}.$ (ii) If $u\in z^{-1}D\Omega^{+}U$
then $u=z^{-1}Dw$ for some $w\in\Omega^{+}U,$ and we have, using
the definitions of $\bar{v}^{-1}$ and of $\bar{\phi},\bar{v}^{-1}j^{+}\underline{u}=\bar{v}^{-1}z^{-1}Dw=\left(\bar{l}^{-1}-\bar{\phi}\right)z^{-1}Dw=$
$z^{-1}\left(\bar{l}^{-1}D-N\right)w.$ Now, in view of (7.5) the
map $\left(\bar{l}^{-1}D-N\right)$ has a strictly polynomial transfer
function so that $z^{-1}\left(\bar{l}^{-1}D-N\right)$ is polynomial.
since also $w$ is polynomial it follows that $z^{-1}\left(\bar{l}^{-1}D-N\right)w\in\Omega^{+}U,$
whence $u\in\operatorname{ker}\pi^{+}\bar{v}^{-1}j^{+}$ as claimed.
$\square$

Proof of Theorem $7.2.$ If $\bar{l}$ is a bicausal precompensator
for $\bar{f}$ and $(\bar{v},\bar{g})$ is a representation of $\bar{l}$,
then $\bar{l}=(I+\bar{g}\cdot\bar{f})^{-1}\bar{v},$ whence $\bar{l}^{-1}=\bar{v}^{-1}+\bar{v}^{-1}\cdot\bar{g}\cdot\bar{f}=\bar{v}^{-1}+\bar{\rho}\cdot\bar{f}$
where the map $\bar{\rho}=\bar{v}^{-1}\bar{g}$ is clearly also causal.
By Lemma $7.4,\bar{v}$ and $\bar{v}^{-1}$ have the same reachability
indices. Hence the theorem will be proved if we can show that $\bar{l}^{-1}$
can be represented as 
\[
\bar{l}^{-1}=\bar{v}^{-1}+\bar{\phi}
\]
satisfying the following requirements: (a) $\bar{v}^{-1}:\Lambda U\rightarrow\Lambda U$
is a bicausal $\Lambda\mathrm{K}$ -linear map such that its reachability
indices $\sigma_{i}$ satisfy $\sigma_{i}\leqq\nu_{i},i=1,\cdots,m.$
(b) The $\Lambda K$ -linear map $\bar{\phi}:\Lambda U\rightarrow\Lambda U$
is strictly causal and can be represented as $\bar{\phi}=\bar{\rho}\cdot\bar{f}$
for some causal $\Lambda K$ -linear map $\bar{\rho}:\Lambda Y\rightarrow\Lambda U$.
As we see below, the maps $\bar{\phi}$ and $\bar{v}^{-1}$ as defined
in (7.6) and (7.7) satisfy the required conditions. Indeed, Lemma
$7.8(\mathrm{i})$ combined with Theorem 5.2 implies that $\bar{\phi}=\bar{p}\cdot\bar{f}$
for some causal $\bar{\rho}.$ since $\bar{f}$ is strictly causal
by definition, it follows that so also is $\bar{\phi}$. Hence condition
(b) above holds. To see that (a) is also satisfied note first that
the difference between a bicausal $\Lambda K$ -linear map and a strictly
causal one is bicausal (see e.g. Corollary 2.11 ). Hence the map $\bar{v}^{-1}$
is bicausal. Now Lemma 7.8 (ii) implies the requirement on the reachability
indices since, in particular, it implies that $\bar{v}^{-1}$ can
be realized with state space $\Omega^{+}U/z^{-1}D\Omega^{+}U$ whose
reachability indices are the column degrees of $z^{-1}D.$ (The reader
is referred to Hautus and Heymann {[}1978{]} for relevant details
on the problem of realization.)$\square$

While Theorem 7.2 gives an upper bound on the required dynamic order
of precompensator remainders, it has been, so far, seen only in the
nonlatent case that this bound is tight. It is clear that in general,
except in the case of nonlatent i/o maps, the maximal required order
of precompensator remainders depends not only on the i/o map $\bar{f}$
but also on the specific precompensator $\bar{l}$ under consideration.
It turns out that the bound of Theorem 7.2 is tight, however, in the
following sense: There always exist bicausal isomorphisms $\bar{l}$
for which all precompensator remainders satisfy the condition that
$n=\sum_{i=1}^{m}\sigma_{i}\geqq\sum_{i=1}^{m}\nu_{i},$ where $n$
is the minimal state space dimension and the $\sigma_{i}$ are reachability
indices of the precompensator remainder, and the $\nu_{i}$ are the
latency indices of the i/o, map $\bar{f}$.
\begin{thm*}
7.9. Let $\bar{f}:\Lambda U\rightarrow\Lambda Y$ be an injective
linear i/o map with latency indices $\nu_{1},\cdots,\nu_{m}.$ There
exists a $\Lambda K$ -linear bicausal isomorphism $\bar{l}:\Lambda\dot{U}\rightarrow\Lambda U$
such that the following holds: If $(\bar{v},\bar{g})$ is any representation
of $\bar{l}$ and if $\sigma_{1},\cdots,\sigma_{m}$ are the reachability
indices of the precompensator remainder $\bar{v},$ then $\sum_{i=1}^{m}\sigma_{i}\geqq\sum_{i=1}^{m}\nu_{i}$.
\end{thm*}
Proof. Let $d_{1},\cdots,d_{m}$ be a proper strictly polynomial basis
for $\ker$. $\pi^{-}\bar{f}$ and write $\ker$ $\pi^{-}\bar{f}=D\Omega^{-}U$
where $D=\left[d_{1},\cdots,d_{m}\right].$ Then the matrix $D_{1}:=z^{-1}D$
is also polynomial and $D_{1}^{-1}$ is causal (see Proposition 6.18
). Below we shall use the same notation interchangeably for matrices
and their associated $\Lambda K$ -linear maps. Let $L:\Lambda U\rightarrow\Lambda U$
be any static $\Lambda K$ -linear map such that $L+D_{1}^{-1}$ is
bicausal. Consider the bicausal pre- compensator $\bar{l}:=\left(L+D_{1}^{-1}\right)^{-1}$.
If $\bar{v}$ is any precompensator remainder for $\bar{l}$, then
$\bar{v}^{-1}=\bar{l}^{-1}-\bar{\rho}\bar{f}=L+D_{1}^{-1}-\bar{\rho}\bar{f}$
for some causal map $\bar{\rho}.$ By Lemma $7.4,\bar{v}$ has the
same reachability indices as $\bar{v}^{-1}$ and the latter has the
same reachability indices as $D_{1}^{-1}-\bar{\rho}\bar{f}$ Now,
we have 
\[
D_{1}^{-1}-\bar{\rho}\cdot\bar{f}=\left(I-\bar{\rho}\cdot\bar{f}\cdot D_{1}\right)D_{1}^{-1}=\bar{I}^{*}\cdot D_{1}^{-1}
\]
where $\bar{l}^{*}=I-\bar{\rho}\cdot\bar{f}\cdot D_{1}$ is bicausal
because the composite $\bar{f}\cdot D_{1}$ is strictly causal, the
latter following since $\ker$ $\pi^{-}\bar{f}\cdot D_{1}=D_{1}^{-1}$
$\ker$ $\pi^{-}\bar{f}=D_{1}^{-1}\left(zD_{1}\right)\Omega^{-}U=z\Omega^{-}U.$
Let $\bar{l}^{*}D_{1}^{-1}=P\cdot Q^{-1}$ be a coprime fraction representation
of $\bar{l}^{*}\cdot D_{1}^{-1}($ see, e.g., Heymann {[}1972{]} or
Hautus and Heymann {[}1978{]} . Then clearly $P$ is nonsingular,
and computing determinantal degrees gives us (because $\bar{l}^{*}$
is bicausal) that 
\[
n:=\operatorname{deg}\operatorname{det}Q=\operatorname{deg}\operatorname{det}P+\operatorname{deg}\operatorname{det}D_{1}\geqq\operatorname{deg}\operatorname{det}D_{1}
\]
since $n$ equals the sum of the reachability indices of the i/o map
$P\cdot Q^{-1}$ the proof is complete.$\square$
\begin{note*}
(added in proof). The reader is also referred to Emre and Hautus {[}1980{]}
, where certain solvability conditions for rational matrix equations
are given that are related to the causal factorization problem.
\end{note*}

\section*{REFERENCES}

F. M. BRASH and J. B. PEARSON {[}1970{]}, Pole placement using dynamic
compensators, IEEE Trans. Automat. Control, AC-15, pp. 34-43.

A. E. EC\textcyr{\CYRK\CYRV\CYRE}RG, JR. {[}1974{]}, A characterization
of linear systems via polynomial matrices and module theory MIT Electronic
Systems Laboratory Rep. ESL-R-528, Mass. Inst. of Tech., Cambridge,
MA.

E. EMRE and M. L. J. HAUTUS {[}1980{]}, A polynomial characterization
of ( $(t,\mathscr{P})$ )-invariant and reachability subspaces, this
Journal, $18,$ pp. $420-436.$

P. L. FALB and W. A. WOLOVICH {[}1967{]}, Decoupling in the design
and synthesis of multivariable control systems, IEEE Trans. Automat.
Control, AC- $12,$ pp. $651-659$.

G. D. FORNEY, JR. {[}1975{]}, Minimal bases of rational vector spaces,
with applications to multivariable linear systems$,$ SIAM J. Control,
13, pp. 493-520.

P. A. FUHRMANN {[}1976{]}, Algebraic system theory: an analyst's point
of view, J. Franklin Inst., 301, pp $521-540$

---{[}1979{]}, Linear feedback via polynomial models, Int. J. Control,
to appear.

E. G. GILBERT {[}1969{]}, The decoupling of multivariable systems
by state feedback, SIAM J. Control, 7, pp. $50-64$.

W. H. GREUB {[}1967{]}, Linear Algebra, 3rd edition, Springer Verlag,
Berlin.

M. L. J. HAUTUS and M. HEYMANN {[}1978{]}, Linear feedback-an algebraic
approach, this Journal, 16, pp. $83-105$.

M. HEYMANN {[}1968{]}, Comments on pole assignment in multi-input
controllable linear systems, IEEE Trans. Automat. Control, AC-13,
pp. 748-749.

M. HEYMANN {[}1972{]}, Structure and realization problems in the theory
of dynamical systems, Lecture Notes, International Center for Mechanical
Sciences, Udine, Italy; also Springer-Verlag, New York, 1975.

R. E. KALMAN, P.L. FALB AND M. A. ARBIB {[}1969{]}, Topics in mathematical
system theory, McGraw Hill, New York.

D. G. LUENBERGER {[}1966{]}, Observers for multivariable systems,
IEEE Trans. Automat. Control, Ac-11, pp. $190-197$.

C. C. MACDUFFEE {[}1934{]}, The Theory of Matrices, Chelsea, New York.

A. S. MORSE $[1975],$ System invariants under feedback and cascade
control, Proceedings of the conference on mathematical systems theory,
Udine, Italy, pp. $61-74;$ Lecture Notes in Economics and Mathematical
Systems $131,$ Springer Verlag, Berlin.

A. S. MORSE and W. M. WONHAM {[}1970{]}, Decoupling and pole assignment
by dynamic compensation, SIAM J. Control, $8,$ pp. $317-337$.

H. F. MÜNZER and D. PRÄTZEL-WOLTERS {[}1979a{]}, Minimal bases of
polynomial modules, structural indices and Brunovsky-transformations,
Int. J. Control, 30, pp. 291-318. 

---$[1979b],$ Geometric and moduletheoretic approach to linear systems,
Part 1: basic categories and functors, Proceedings of the Delft Conference
on Systems and Networks, July. 

---$[1979\mathrm{c}]$ Geometric and moduletheoretic approach to
linear systems, Part 2: moduletheoretic characterization of reachability
subspaces, Internal report, Universität Bremen, Bremen, Germany.

H. H. ROSENGROCK $[1970],$ State space and multivariable theory,
Nelson, London.

J. D. SIMON and S. K. MITTER {[}1968{]} , $A$ theory of modal control,
Information and Control, $13,$ pp. $316-353$.

W. A. WOLOVICH {[}1974{]}, Linear multivariable systems, Applied Mathematical
Sciences Series, 11, Springer-Verlag, New York.

W. M. WONHAM {[}1967{]}, On pole assignment in multi-input controllable
linear systems, IEEE Trans. Automat. Control. AC-12, pp. 660-665.

---{[}1979{]} Linear Multivariable Control: A Geometric Approach,
2nd ed., Springer-Verlag, New York.

W. M. WONHAM and A. S. Morse {[}1970{]}, Decoupling and pole assignment
in linear multivariable systems: A geometric approach, SIAM J. Control,
$8,$ pp. $1-18$.

B. F. WYMAN {[}1972{]}, Linear systems over commutative rings, Lecture
notes, Stanford Univ., Stanford, CA.
\end{document}